\documentclass[12pt]{article}%
\usepackage{amsfonts}
\usepackage{amsmath}
\usepackage{bm}
\usepackage{hyperref}
\usepackage{graphicx}
\usepackage{amssymb}
\usepackage[doublespacing]{setspace}
\usepackage{natbib}
\usepackage{geometry}
\usepackage{sectsty}
\usepackage[font=small]{caption}%
\providecommand{\U}[1]{\protect\rule{.1in}{.1in}}
\providecommand{\U}[1]{\protect\rule{.1in}{.1in}}
\numberwithin{equation}{section}
\newtheorem{theorem}{Theorem}

\newtheorem{corollary}[theorem]{Corollary}

\newtheorem{lemma}{Lemma}

\newtheorem{proposition}{Proposition}

\ifx\pdfoutput\relax\let\pdfoutput=\undefined\fi
\newcount\msipdfoutput
\ifx\pdfoutput\undefined\else
\ifcase\pdfoutput\else
\ifx\paperwidth\undefined\else
\ifdim\paperheight=0pt\relax\else\pdfpageheight\paperheight\fi
\ifdim\paperwidth=0pt\relax\else\pdfpagewidth\paperwidth\fi
\fi\fi\fi
\begin{document}

\title{\textbf{Minimax regret treatment rules with finite samples when a quantile is
the object of interest}\thanks{We would like to thank the Editor Peter
Phillips, the CoEditor Xiaoxia Shi, and an anonymous referee for many helpful
suggestions. We also would like to thank Antonio Galvao, Kei Hirano, seminar
participants at (in chronological order) UC Davis, UC Riverside, Colorado
Boulder, SUNY Stony Brook, McGill, M\"{u}nchen, Konstanz, Frankfurt,
Heidelberg, Regensburg, NTU, SMU, NUS, Macquarie University, University of
Sydney, UNSW, Brisbane, UCL, Exeter, Cambridge, Manchester, Oxford, conference
participants at the New York Camp Econometrics, Lake Placid, and the Rochester
conference in Econometrics for helpful comments.}}
\author{$%
\begin{array}
[c]{c}%
\text{Patrik Guggenberger}\\
\text{Department of Economics}\\
\text{Pennsylvania State University}%
\end{array}
$
\and $%
\begin{array}
[c]{c}%
\text{Nihal Mehta}\\
\text{Keystone}\\
\end{array}
$
\and $%
\begin{array}
[c]{c}%
\text{Nikita Pavlov}\\
\text{Department of Economics}\\
\text{Pennsylvania State University}%
\end{array}
$}
\date{First Version: July, 2023; Revised: \today}
\maketitle

\begin{abstract}
Consider a setup in which a decision maker is informed about the population by
a finite sample and based on that sample has to decide whether or not to apply
a certain treatment. We work out finite sample minimax regret treatment rules
under various sampling schemes when outcomes are restricted onto the unit
interval. In contrast to Stoye (2009) where the focus is on maximization of
expected utility the focus here is instead on a particular quantile of the
outcome distribution. We find that in the case where the sample consists of a
fixed number of untreated and a fixed number of treated units, any treatment
rule is minimax regret optimal. The same is true in the case of random
treatment assignment in the sample with any assignment probability and in the
case of testing an innovation when the known quantile of the untreated
population equals 1/2. However if the known quantile exceeds 1/2 then never
treating is the unique optimal rule and if it is smaller than 1/2 always
treating is optimal. We also consider the case where a covariate is included.



\end{abstract}

\section{Introduction\label{Introduction}}

Consider a setup in which a decision maker (DM) is informed about the
population by a finite sample drawn from the population and based on that
sample has to decide whether or not to apply a certain treatment or whether to
randomize treatment assignment. The DM could be a policymaker who applies a
treatment to the entire population or a person who is applying the treatment
to an individual (e.g., herself). As examples of the latter setup, think of an
individual who books a hotel accommodation on an internet platform after
observing a certain number of ratings or a medical doctor who picks a
treatment for a patient after observing a certain number of outcomes on the treatments.

This paper is adding to a short but growing literature on finding treatment
rules, i.e. measurable mappings from the sample to the unit interval, that
have finite-sample optimality properties. In most of the literature, the focus
is on the expected outcome under the chosen treatment rule.\footnote{See e.g.
Manski (2004), Manski and Tetenov (2007), Stoye (2007, 2009, 2012), Tetenov
(2012), Masten (2023), Montiel Olea, Qiu, and Stoye (2023), Yata (2023),
Kitagawa, Lee, and Qiu (2024), Chen and Guggenberger (2025), and additional
references in these papers. Hirano and Porter (2009), Kitagawa and Tetenov
(2018), and Christensen, Moon, and Schorfheide (2023) and many other
references therein also use the minimax regret criterion but consider an
asymptotic, rather than a finite sample, framework. This literature is
inspired by the classical work of Wald (1950). In a recent paper Manski and
Tetenov (2023) study several potential features of the state-dependent
distribution of loss (rather than just its expectation) that a decision rule
generates across potential samples.} Given that typically there is no
treatment rule that is uniformly best over all possible joint distributions
for $(Y_{0},Y_{1}),$ where $Y_{0}$ and $Y_{1}$ denote random variables for
outcomes without and with treatment, respectively, one has to resort to other
criteria for optimality. One option is to consider a prior over the space of
joint distributions and maximize expected outcome for this particular prior.
Another option is to focus on admissible treatment rules but that criterion
typically does not single out an individual treatment rule, see Manski and
Tetenov (2023) and Montiel Olea, Qiu, and Stoye (2023). Alternatively, one
might consider finding a treatment rule that maximizes minimal expected
outcome where the minimum is taken over all joint distributions of
$(Y_{0},Y_{1}).$ However, if there exists a distribution that assigns the
minimal possible values in the shared domains of $Y_{0}$ and $Y_{1}$ with
probability one then any treatment rule is going to be optimal according to
this criterion and therefore also the \textquotedblleft
max-min\textquotedblright\ approach is not pinning down a unique rule. For
that reason, instead, the so-called \textquotedblleft minimax
regret\textquotedblright\ criterion is often adopted that determines treatment
rules that minimize the maximal regret where regret is defined as the
difference between the largest expected outcome that could be achieved for any
treatment rule and the expected outcome under the treatment rule under
consideration, and the maximum is taken with respect to all possible
distributions for $(Y_{0},Y_{1}).$ Stoye (2009) derives minimax regret rules
in finite samples for the case of two treatments under various sampling
schemes, namely matched pairs, random sampling, and testing an innovation, and
furthermore provides near-uniqueness results.

In this project we are interested in a setup where rather than expected
outcome, the DM is concerned about the $\alpha$-quantile of the outcome (for a
given $\alpha\in\lbrack0,1]$). Wang et al. (2018) consider robust estimation
of the quantile-optimal treatment regime and provide arguments as to why
focusing on a quantile rather than the mean may be sensible in certain
applications. In fact, in many applications the tail of the outcome
distribution is at the center of interest. For example, when evaluating job
training programs to improve earnings the focus is often for earnings in the
lower tail and likewise in survival analysis (e.g., survival time of cancer
patients) the lower tail is of key importance. Wang et al. (2018, Section 2)
provide numerical evidence where the mean-optimal treatment is detrimental for
patients in the lower tail. In Economics, one might be interested in median
income (case $\alpha=1/2)$ or a minimal education achievement for school kids,
\textquotedblleft no child left behind\textquotedblright\ (case $\alpha=0$).
See Manski (1988) who studies the \textquotedblleft quantile
utility\textquotedblright\ model whose predictions, unlike the expected
utility model, are invariant under ordinal transformations of utility.
Subsequently, Rostek (2010) axiomatizes quantile preference and, recently,
Manski and Tetenov (2023) suggest considering various deviations from mean
loss including quantiles. Also see Chambers (2009).\footnote{Related (but in a
non-finite-sample setup) Qi, Cui, Liu, and Pang (2019) and Qi, Pang, and Liu
(2023) consider optimal decision rules based on the conditional value at risk
(CVaR) measure. Quantile preferences have been attracting growing interest in
the literature, e.g. De Castro, Galvao, and Ota (2026) consider a model in
which an economic agent maximizes the discounted value of a stream of future
$\alpha$-quantile utilities.}

As the main contribution of this paper, we derive minimax regret treatment
rules $\delta$ in finite samples when an $\alpha$-quantile of the outcome
distribution is the focus of interest and outcomes $Y_{0}$ and $Y_{1}$ take
values in the unit interval. Somewhat surprisingly, we show that under various
sampling schemes \emph{all} treatment rules are minimax regret, namely in the
case i) when the sample consists of a fixed number of treated and a fixed
number of untreated units (that is, unbalanced panels are allowed for) and in
the case ii) under random assignment with arbitrary treatment assignment
probability in the sample equal to $p\in(0,1).$ In both cases i) and ii)
maximal regret equals 1 for any treatment rule.

On the other hand, in the case iii) \textquotedblleft testing an
innovation\textquotedblright, that is, the case where only data on $Y_{1}$ is
observed and the $\alpha$-quantile $q_{s,\alpha}(Y_{0})$ of $Y_{0}$ is known,
where the subscript $s$ denotes the joint distribution of $(Y_{0},Y_{1})$, if
$q_{s,\alpha}(Y_{0})>1/2$ then $\delta\equiv0$ is the unique minimax rule, if
$q_{s,\alpha}(Y_{0})<1/2$ then $\delta\equiv1$ is a minimax rule, and finally,
if $q_{s,\alpha}(Y_{0})=1/2$ then any treatment rule $\delta$ is minimax
regret; in each case, for minimax treatment rules $\delta$ we obtain the
formula $\min\{q_{s,\alpha}(Y_{0}),1-q_{s,\alpha}(Y_{0})\}$ for the maximal regret.

Obviously, like for the case where expected outcomes are the focus, also here
the max-min criterion is not informative. Namely, as long as the joint
distribution of $(Y_{0},Y_{1})$ can be chosen such that $q_{s,\alpha}%
(Y_{0})=q_{s,\alpha}(Y_{1})=0,$ that is, the worst possible outcome under the
given restriction that $Y_{0},Y_{1}\in\lbrack0,1],$ any treatment rule would
be max-min optimal. In contrast though, to the case where expected outcomes
are the focus, for quantiles also minimax regret is not informative for
sampling designs i) and ii) and also for iii) when $q_{s,\alpha}(Y_{0})=1/2.$

The typical strategy in the extant literature for determining minimax regret
rules when the focus is on expected outcomes is via a Nash equilibrium
approach in a fictitious zero sum game in which the DM plays against an
antagonistic nature whose payoff equals regret. To establish that a particular
treatment rule $\delta$ is minimax regret one attempts to guess a
\textquotedblleft state of nature\textquotedblright\ $s$ , allowed to be mixed
strategy (over a finite number of states), called a \emph{least favorable
distribution}, for which the pair $(\delta,s)$ constitutes a Nash equilibrium.
Existence of such a pair $(\delta,s)$ implies that $\delta$ is indeed a
minimax regret rule, see for example Berger (1985), Stoye (2009), and Chen and
Guggenberger (2025). Often, in a first step, one restricts outcomes to be
Bernoulli and finds a minimax rule in this simplified setup and then, in a
second step, uses the so-called coarsening approach to tackle the general
case, see e.g., Cucconi (1968), Gupta and Hande (1992), and Schlag (2003,
2006). When nature picks a state of the world trying to inflict high regret it
faces the trade-off that on the one hand a high differential between expected
outcomes with and without treatment is needed but on the other hand the more
different the distributions of $Y_{0}$ and $Y_{1}$ are the easier the DM can
tell them apart using the sample information.

The proof structure for the main results in this paper differs from the one
just described. Namely for cases i)\ and ii) we show that for any treatment
rule $\delta$ one can find a state of nature $s=s_{\delta}$ such that $s$
inflicts the highest possible regret, namely 1. That insight is sufficient to
establish that all rules are minimax regret and that maximal regret equals 1
for all treatment rules. It is noteworthy and remarkable that, despite the
trade-off just described, nature is powerful enough to inflict maximal regret
on the DM. Even more surprisingly, the conclusion under i) and ii) continue to
be true even if nature is restricted to only Bernoulli distributions. The
proof for part iii) relies on the main insight that if $\delta\neq0$ then
there exists a state of nature such that regret equals the known $\alpha
$-quantile of $Y_{0}.$ For that to be true it is sufficient for nature to have
discrete distributions, supported on $N+1$ points, at its disposal.

We start off with the case with no covariates and then allow for a
discrete-valued covariate in the model. In the latter case, a treatment rule
maps a sample onto treatment probabilities for each of the $K$ possible
outcomes of the covariate. Not surprisingly, given the results without a
covariate, we find that in cases i), ii), and case iii) with known quantile
equal to 1/2, again, each treatment rule is minimax regret, while in case iii)
when the known quantile is different from 1/2, no-data rules are minimax regret.

We include a small finite sample simulation study in the case of
\textquotedblleft testing an innovation\textquotedblright\ where we simulate
regret of various treatment rules, namely the empirical success rule and
several no-data rules. The study corroborates our theoretical findings about
minimax regret treatment rules and the formulas for maximal regret that we derive.

The remainder of the paper is organized as follows. Section 2 introduces the
theoretical setup for our notion of a minimax regret rule with quantiles and
contains analytical results when no covariates are included. In Subsection 2.1
we present various approaches for how quantiles could be incorporated into the
Waldean framework of statistical decision theory and juxtapose them with our
approach. Subsection 2.2 derives minimax regret treatment rules under various
sampling schemes and also provides a brief discussion of minimax regret
treatment rules when certain restrictions are imposed on the states of nature.
Subsection 2.3 simulates the performance of various treatment rules in finite
samples in the case of testing an innovation. Finally, Section 3 derives
minimax regret treatment rules in the case where covariates are included in
the model. All proofs are given in the Appendix.

\section{Theoretical Setup Without Covariates\label{setup}}

A decision maker has to decide whether or not to assign treatment after being
informed about the population by a finite sample.\footnote{Alternatively,
rather than framing the options as "treatment" and "no treatment", one could
frame the setup as a choice between two treatments.} The setup is very similar
to the one in Stoye (2009) with one key modification. Namely, rather than
focusing on mean outcomes, here we are concerned with the $\alpha$-quantile of
the outcome distribution.

For most parts of the paper potential outcomes $Y_{0}$ and $Y_{1}$ for
untreated/treated individuals are restricted to the unit interval
\begin{equation}
Y_{0},Y_{1}\in S:=[0,1], \label{outcome space}%
\end{equation}
where $S$ is assumed known to the DM. At first, different members of the
population are all identical to the decision maker. Later, we will consider
the case where a covariate is included and treatment assignment can be made
conditional on the realization of the covariate.

By $\mathbb{S}$ we denote the set of \textquotedblleft states of the
world\textquotedblright\ $s$ where an $s$ denotes a possible joint probability
distribution over the potential outcomes for $Y_{0}$ and $Y_{1}.$ Unless
otherwise stated $\mathbb{S}$ is unrestricted and contains all possible joint
distributions for $Y_{0}$ and $Y_{1}$ on $S^{2}.$ Upon observing a sample
$w_{N}=(t,y)$ of size $N$ of treatment statuses $t=(t_{1},...,t_{N})$ and
outcomes $y=(y_{1},...,y_{N})$ where the $i$-th component of $y,$ $y_{i},$ is
an independent realization of $Y_{t_{i}}$ and $t_{i}$ denotes the treatment
received by individual $i,$ the task for the decision maker is to choose
\begin{equation}
\delta(w_{N})\in\lbrack0,1], \label{treatment rule}%
\end{equation}
which denotes the probability with which treatment is assigned.\footnote{We do
not index $t$ and $y$ by $N$ because it would make the notation too
cumbersome.
\par
{}} Namely, which treatment the DM assigns is determined as an independent
draw of the Bernoulli random variable $B=B(\delta(w_{N}))\in\{0,1\}$ that
equals $1$ with probability $\delta(w_{N})$ and is assumed to be independent
of all other random objects. Therefore, the setup here allows for randomized
treatment rules.

Let $\alpha\in\lbrack0,1].$ Given a state of the world $s$ and a statistical
treatment rule $\delta$ the objective function for the DM is
\begin{equation}
u(\delta,s)=q_{s,\alpha}(Y_{B(\delta(w_{N}))}),
\label{objective of decision maker}%
\end{equation}
where $Y_{B(\delta(w_{N}))}$ denotes random outcomes generated when the
treatment rule $\delta$ is used, and $q_{s,\alpha}(Y_{B(\delta(w_{N}))})$
denotes the $\alpha$-quantile of $Y_{B(\delta(w_{N}))}$ when the state of the
world is $s\in\mathbb{S}.$ In particular, when the treatment rule
$\delta(w_{N})$ equals 0 (or 1) then with probability 1 $Y_{B(\delta(w_{N}))}$
equals $Y_{0}$ (or $Y_{1}$).

By definition, an $\alpha$-quantile of a scalar valued random variable $X$
with domain $D$ is any number $q\in D$ that satisfies
\begin{equation}
P(X\leq q)\geq\alpha\text{ and }P(X\geq q)\geq1-\alpha.
\label{quantile definition}%
\end{equation}
Clearly then, any $\alpha$-quantile of $Y_{0},$ $Y_{1}$ and $Y_{B(\delta
(w_{N}))}$ is an element of [0,1]. In general, this definition does not lead
to a unique $\alpha$-quantile. The definition allows for the case where a
quantile has a non-zero point mass. To be explicit in cases where there is
non-uniqueness, we make the following definition that hinges on a choice
$r\in\lbrack0,1].$

\textbf{Definition }$\alpha$\textbf{-quantile:} Let $Q$ denote the set of all
$\alpha$-quantiles $q$. For $\alpha\in(0,1)$ we define the $\alpha$-quantile
as
\begin{equation}
r\sup Q+(1-r)\inf Q. \label{definition of the alpha quantile}%
\end{equation}
When $\alpha=0$ we use $\sup Q$ (that is, we use $r=1$) and when $\alpha=1$ we
use $\inf Q$ (that is, we use $r=0).\medskip$

From now on, by $q_{s,\alpha}(Y_{B(\delta(w_{N}))})$ we denote the $\alpha
$-quantile of $Y_{B(\delta(w_{N}))}$ when the state of the world is $s.$ For
simplicity of notation, unless needed for clarity, we do not index that
expression by $r.$

By definition, regret of the treatment rule $\delta$ for a given distribution
$s$ of $(Y_{0},Y_{1})$ equals%
\begin{equation}
R(\delta,s)=\sup_{d\in\mathbb{D}}u(d,s)-u(\delta,s),\label{regret}%
\end{equation}
where $\mathbb{D}$ denotes the set of all possible treatment rules. In words,
regret equals the difference between the highest $\alpha$-quantile that could
have been achieved for any treatment rule $d\in\mathbb{D}$ and the $\alpha
$-quantile obtained for the particular treatment rule $\delta$ for the given
state of nature $s.$ If $q_{s,\alpha}(Y_{1})\geq q_{s,\alpha}(Y_{0})$
($q_{s,\alpha}(Y_{1})<q_{s,\alpha}(Y_{0}))$ and $d^{\ast}=1$ ($d^{\ast}=0$) is
an element of $\mathbb{D}$ then $\sup_{d\in\mathbb{D}}u(d,s)$ is taken on by
$d^{\ast}$ and equals $q_{s,\alpha}(Y_{1})$ ($q_{s,\alpha}(Y_{0})$), see Lemma
3(ii)\textbf{ }in the Appendix. If restrictions are imposed on $\mathbb{D}$
then $\sup_{d\in\mathbb{D}}u(d,s)$ may not be taken on by any element in
$\mathbb{D}\mathfrak{.}$ Again, without restrictions, $1(q_{s,\alpha}%
(Y_{1})\geq q_{s,\alpha}(Y_{0}))$ is an infeasible optimal rule$,$ where
$1(\cdot)$ denotes the indicator function.

In this paper, we focus on minimax regret treatment rules. By definition, if
it exists, such a rule satisfies%
\begin{equation}
\delta^{\ast}\in\arg\min_{\delta\in\mathbb{D}}\sup_{s\in\mathbb{S}}%
R(\delta,s). \label{minimax rule}%
\end{equation}

In contrast, a maximin treatment rule, if it exists, satisfies
\begin{equation}
\delta^{+}\in\arg\max_{\delta\in\mathbb{D}}\inf_{s\in\mathbb{S}}u(\delta,s).
\label{maximin}%
\end{equation}
As discussed already elsewhere (see e.g., Manski (2004), Stoye (2009)) the
maximin criterion may lead to the uninformative result that \emph{all}
$\delta\in\mathbb{D}$ are maximizers. That occurs e.g., if for a particular
$s^{+}\in\mathbb{S}\mathfrak{,}$ $u(\delta,s^{+})$ does not depend on $\delta$
and takes on its smallest possible value, $u(\delta,s^{+})=\inf_{s\in
\mathbb{S}}u(\delta,s).$ That situation also occurs in our setup where the
objective is to maximize the quantile of the outcome distribution, namely when
$s^{+}\in\mathbb{S}$ is chosen such that $q_{s,\alpha}(Y_{0})=q_{s,\alpha
}(Y_{1})=0.$ In contrast to the setup where expected outcomes are the
objective, as we will establish next, it turns out that for quantiles,
depending on the particular sampling design, also the minimax regret criterion
may be uninformative.

\subsection{Waldean statistical decision theory and
quantiles\label{Waldean DT}}

In this subsection\footnote{This section is inspired by the constructive
comments of an anonymous referee.} we further discuss the proposed criterion
introduced in (\ref{objective of decision maker}) and juxtapose it with other
possible approaches of how quantiles could be incorporated into a Waldean
framework of statistical decision theory.

In a series of pathbreaking contributions Wald (1945, 1947, 1950) introduced a
framework for statistical decision theory whose main components are a
statistical model, an action space, statistical decision rules, a welfare and
an expected welfare function, and finally an evaluation criterion. Instead of
welfare and expected welfare the presentation could be framed in terms of loss
(defined as negative welfare) and risk (defined as expected loss).\ See Hirano
(2025) for a comprehensive review that also includes contemporary developments.

In the particular context considered here where the DM\ needs to pick
treatment 0 or 1 after observing the i.i.d. sample $w_{N}$ a minimax regret
rule in the Waldean formulation solves%
\begin{equation}
\arg\min_{\delta\in\mathbb{D}}\max_{s\in\mathbb{S}}[\max\{\mu_{0},\mu
_{1}\}-E_{s}Y_{B(\delta(w_{N}))}], \label{Waldean 0}%
\end{equation}
where $\mu_{t}=E_{s}Y_{t}$ for $t=0,1$ and $E_{s}$ denotes the expectation
operator under $s.$ It is easily shown that $s$ matters only via $(\mu_{0}%
,\mu_{1}).$ By the law of iterated expectations,%
\begin{align}
E_{s}Y_{B(\delta(w_{N}))}  &  =E_{w_{N}}E_{s}(Y_{B(\delta(w_{N}))}%
|w_{N})\nonumber\\
&  =\mu_{0}E_{w_{N}}(1-\delta(w_{N}))+\mu_{1}E_{w_{N}}\delta(w_{N}),
\label{waldean 1}%
\end{align}
overall uncertainty can be separated into \emph{sampling uncertainty }through
$w_{N},$ (potential) randomness through the \emph{treatment assignment} $B,$
and randomness through the \emph{potential outcome variables} $(Y_{0},Y_{1}%
)$\emph{;} \emph{risk} is defined as the average loss over sampling uncertainty.

To adapt the Waldean framework to one that is based on the notion of $\alpha
$-quantile rather than expectation, we suggest replacing expectations by
$\alpha$-quantiles in the formulation (\ref{Waldean 0}), that is, we suggest
solving (\ref{minimax rule}) which is%
\begin{equation}
\arg\min_{\delta\in\mathbb{D}}\max_{s\in\mathbb{S}}[\max\{q_{s,\alpha}%
(Y_{0}),q_{s,\alpha}(Y_{1})\}-q_{s,\alpha}(Y_{B(\delta(w_{N}))})].
\label{our approach again}%
\end{equation}
Given there is no equivalent to the law of iterated expectations for
quantiles, by doing so, one loses the separation of the regret criterion into
loss without sampling uncertainty and sampling uncertainty. Our proposed
criterion aggregates joint uncertainty (sampling, treatment assignment, and
outcome) before taking the quantile.\footnote{Note that $q_{s,\alpha
}(Y_{B(\delta(w_{N}))}|w_{N})$ does not in general equal $\delta
(w_{N})q_{s,\alpha}(Y_{1})+(1-\delta(w_{N}))q_{s,\alpha}(Y_{0})$ in cases
where $\delta(w_{N})\in(0,1).$} \smallskip

Instead, to maintain such separation, one could first define regret loss given
a realized sample $w_{N}$ as%
\begin{equation}
L(\delta(w_{N}),s)=\max\{q_{s,\alpha}(Y_{0}),q_{s,\alpha}(Y_{1})\}-q_{s,\alpha
}(Y_{B(\delta(w_{N}))}|w_{N}) \label{cond loss}%
\end{equation}
and then define an alternative regret to (\ref{regret}) as the average of
regret loss over the sampling distribution $w_{N}$%
\begin{equation}
R^{\prime}(\delta,s)=\max\{q_{s,\alpha}(Y_{0}),q_{s,\alpha}(Y_{1})\}-E_{w_{N}%
}q_{s,\alpha}(Y_{B(\delta(w_{N}))}|w_{N}). \label{regret risk}%
\end{equation}
The first term in (\ref{cond loss}) is a benchmark attainable if one could
pick the better treatment in terms of quantile outcome without sampling
uncertainty; the second term is the $\alpha$-quantile of $Y_{B(\delta(w_{N}%
))}$ conditional on the realized sample $w_{N}$. This criterion may be
misaligned when the DM cares about tails in the distribution of $q_{s,\alpha
}(Y_{B(\delta(w_{N}))}|w_{N})$ over realized samples rather than the average
over sampling uncertainty. As yet another alternative, one could consider a
regret function defined as the $\beta$-quantile (for some $\beta$ in the unit
interval) of the loss function in (\ref{cond loss}), that is
\begin{equation}
R^{\prime\prime}(\delta,s)=q_{s,\beta}(L(\delta(w_{N}),s)).
\label{regret option 2}%
\end{equation}
By the monotone transform identity of quantiles it follows that
\begin{equation}
R^{\prime\prime}(\delta,s)=\max\{q_{s,\alpha}(Y_{0}),q_{s,\alpha}%
(Y_{1})\}-q_{s,1-\beta}(q_{s,\alpha}(Y_{B(\delta(w_{N}))}|w_{N}))
\label{regret option 2 rewritten}%
\end{equation}
and the interpretation of the criterion is not straightforward.\smallskip

Manski and Tetenov (2023, Section 6.1) suggest an entire class of alternative
approaches that also maintain separation. Starting with an arbitrary loss
function (for instance negative welfare) for a given action by the DM (i.e. in
our setup, a choice of treatment 0 or 1) and DGP\ $s,$ the suggestion is to
consider for example an $\alpha$-quantile of the loss function over the
sampling distribution. In our setup, such an approach could be formulated as
\begin{equation}
\arg\min_{\delta\in\mathbb{D}}\sup_{s\in\mathbb{S}}[\max\{\mu_{0},\mu
_{1}\}-q_{s,\alpha}(\mu_{B(\delta(w_{N}))})]. \label{minmax regret MT}%
\end{equation}

One issue with implementation of the various criteria is that analytical
formulas are not generally available.\footnote{Recently suggested numerical
procedures for the implementation of minimax rules by Aradillas Fern\'{a}ndez
et al. (2025) and Guggenberger and Huang (2025) might be applicable also for
these scenarios.}\smallskip

We do not have a strong opinion about which one of the above criteria is
generally preferable. However, we find our criterion very natural in
situations where a DM applies treatment to an individual (e.g., herself) after
observing the sample. Consider, for example, an individual who books a hotel
accommodation on an internet platform, picking one of two options after
observing a certain number of ratings on each (or picks one of two restaurants
after having observed a number of ratings for each on the internet). The
outcome is the rating the DM assigns to the hotel she booked, which could be
interpreted as a proxy for the welfare the DM received from staying at the
hotel. Another example is a medical doctor who picks one of two treatments for
a patient after observing a certain number of outcomes on the two treatments.

In that type of example, when repeating the exercise, every observed outcome
combines the randomness of the sample, (potential) randomness of treatment
assignment, and the randomness of $(Y_{0},Y_{1}).$ If the DM is concerned
about quantiles of the outcome distribution it seems that the criterion
proposed in (\ref{our approach again}) is a natural choice.

\subsection{Minimax Regret Treatment Rules\label{No covariates}}

Recall that a sample of size $N$ has the general structure $w_{N}=(t,y).$ We
consider three different sample designs, namely:

(i) \emph{Fixed number of untreated/treated units} with $N_{0}\in\{0,...,N\}$
i.i.d. observations of $Y_{0}$ and $N_{1}:=N-N_{0}$ i.i.d. observations of
$Y_{1}$ for some $N\in\mathbb{N}\cup\{0\}$. Here, the notation for a sample
can be simplified by dropping $t.$ We write $w_{N}=(y(0)^{\prime}%
,y(1)^{\prime})^{\prime}\in$ $[0,1]^{N}$ where $y(0)\in\lbrack0,1]^{N_{0}}$
contains the $N_{0}$ observations on untreated units and $y(1)\in
\lbrack0,1]^{N_{1}}$ contains the $N_{1}$ observations on the treated units. A
treatment rule $\delta\in\mathbb{D}$ is then any mapping $[0,1]^{N}%
\rightarrow\lbrack0,1].$ It assigns a treatment probability $\delta(w_{N}%
)\in\lbrack0,1]$ after observing the sample.\smallskip

(ii)\emph{ Random assignment} with $N\in\mathbb{N}\cup\{0\}$ i.i.d.
observations, where in the sample, the treatment probability equals
$p\in(0,1).$\footnote{Note that $p\in\{0,1\}$ leads back to design (i) with
$N_{0}$ or $N_{1}$ equal to $N.$} A treatment rule $\delta\in\mathbb{D}$ is
any mapping $\{0,1\}^{N}\times\lbrack0,1]^{N}\rightarrow\lbrack0,1].$ The rule
$\delta$ assigns a treatment probability $\delta(w_{N})\in\lbrack0,1]$ after
observing a sample $w_{N}=(t,y)$ of treatment statuses $t=(t_{1},...,t_{N})$
and realizations $y=(y_{1},...,y_{N}),$ where the $i$-th component of $y,$
$y_{i},$ is an independent realization of $Y_{t_{i}}$.\smallskip

(iii) \emph{Testing an innovation}, is the case where aspects of the
distribution of $Y_{0}$ are known to the DM; in particular, we assume that the
$\alpha$-quantile of $Y_{0}$ is known (but nature can pick arbitrary
distributions for $Y_{0}$ subject to that restriction). That is, in this case
the set $\mathbb{S}$ consists of all joint distributions $s$ for $(Y_{0}%
,Y_{1})$ with the restriction that the $\alpha$-quantile of the marginal for
$Y_{0}$ equals a certain value $q_{\alpha}(Y_{0}).$ In this case,
$N\in\mathbb{N}\cup\{0\}$ i.i.d. observations of $Y_{1}$ are observed. Here
again the notation for a sample can be simplified. We write $w_{N}%
=y(1)\in\lbrack0,1]^{N}$ where $y(1)$ contains the $N_{1}$ observations of the
treated units. A treatment rule $\delta\in\mathbb{D}$ is any mapping
$[0,1]^{N}\rightarrow\lbrack0,1].$ The rule $\delta$ assigns a treatment
probability $\delta(w_{N})\in\lbrack0,1]$ after observing a sample
$w_{N}=(y_{1,1},...,y_{1,N})$ of $N$ independent realizations $y_{1,i},$
$i=1,...,N$ of $Y_{1}.$\smallskip

The designs above nest the ones in Stoye (2009). In contrast to Stoye (2009),
in design (i) we allow for arbitrary numbers of treated and untreated units
rather than $N/2$ units each as in \textquotedblleft matched
pairs\textquotedblright\ and in design (ii) the treatment probability is any
fixed number $p\in(0,1)$ rather than necessarily $p=.5$.

The following statement provides the analogue to Proposition 1 in Stoye (2009)
and is the main contribution of this paper.

\begin{proposition}
\label{Analogue to Proposition 1 in Stoye 2009}$($i$)$ Consider the case where
the sample consists of a \emph{\textquotedblleft fixed number of treated/
untreated units\textquotedblright.} If $(\alpha\in(0,1)$ and $r\in
\lbrack0,1])$ or $(\alpha=0$ and $r=1)$ then any treatment rule $\delta
\in\mathbb{D}$ is minimax regret and $\max_{s\in\mathbb{S}}R(\delta,s)=1$ for
any $\delta\in\mathbb{D}\mathfrak{.}$ If $(\alpha=1$ and $r=0)$ then exactly
those treatment rules $\delta\in\mathbb{D},$ that are not equal to 1 wp1 and
not equal to 0 wp1, are minimax regret and satisfy $\max_{s\in\mathbb{S}%
}R(\delta,s)=0$.

$($ii$)$ Consider the case where the sample is generated via
\emph{\textquotedblleft random assignment\textquotedblright. }Then the same
statement as in part $($i$)$ holds.

$($iii$)$ In the case of \emph{\textquotedblleft testing an
innovation\textquotedblright} let $(\alpha\in(0,1)$ and $r\in\lbrack0,1])$ or
$(\alpha=0$ and $r=1).$ If the known $\alpha$-quantile of $Y_{0},$ $q_{\alpha
}(Y_{0}),$ equals $1/2$ then any treatment rule $\delta\in\mathbb{D}$ is
minimax regret; if instead $q_{\alpha}(Y_{0})>1/2$ then $\delta^{0}\equiv0$ is
the unique minimax regret rule; if $q_{\alpha}(Y_{0})<1/2$ then $\delta
^{1}\equiv1$ is a minimax regret rule; in each case, for minimax regret
treatment rules $\delta$ we obtain $\max_{s\in\mathbb{S}}R(\delta
,s)=\min\{q_{\alpha}(Y_{0}),1-q_{\alpha}(Y_{0})\}.$ If $(\alpha=1$ and $r=0)$
then exactly those treatment rules $\delta\in\mathbb{D},$ that are not equal
to 1 wp1 and not equal to 0 wp1, are minimax regret and satisfy $\max
_{s\in\mathbb{S}}R(\delta,s)=0$.
\end{proposition}

\textbf{Comments.} 1. Proposition
\ref{Analogue to Proposition 1 in Stoye 2009} establishes that the minimax
regret criterion when applied to $\alpha$-quantiles does not favor data-driven
rules. In fact, in the \emph{\textquotedblleft testing an
innovation\textquotedblright} case data-driven rules are strictly dominated by
$\delta^{0}\equiv0$ when $q_{\alpha}(Y_{0})>1/2$ and weakly dominated by
$\delta^{1}\equiv1$ when $q_{\alpha}(Y_{0})<1/2.$ When $q_{\alpha}(Y_{0})=1/2$
all treatment rules are minimax. This is in stark contrast to the results in
Stoye (2009) about minimax regret treatment rules when the focus is on mean
outcomes. Namely Stoye (2009) shows that e.g., in the case of binary outcomes
where the sample is obtained as a matched pair, to be minimax regret optimal,
the treatment that has more successes in the sample must be chosen with
probability one.\smallskip

2. To provide intuition of the result in (i)-(ii) assume $\alpha\in(0,1)$ and
consider first the simplest case where the sample size $N$ is 0. In that case,
a treatment rule $\delta$ is simply an element in $[0,1]$ that denotes the
probability of assigning treatment. For given $\delta$, the objective for
nature is to find a distribution for $(Y_{0},Y_{1})$ such that the $\alpha
$-quantiles of the marginals have maximal distance (that is distance 1) and
such that the $\alpha$-quantile of $Y_{B(\delta)}$ is zero. Let $\delta>0$
first$.$ Assume nature picks $Y_{0}$ and $Y_{1}$ as independent Bernoulli%
\begin{equation}
P(Y_{1}=0)=1,\text{ }P(Y_{0}=0)=\alpha-\varepsilon\text{ for some }%
\varepsilon\in(0,\alpha] \label{def Y0 Y1}%
\end{equation}
and, consequently, $P(Y_{0}=1)=1-(\alpha-\varepsilon).$ Thus $q_{s,\alpha
}(Y_{1})=0$ and $q_{s,\alpha}(Y_{0})=1$ and%
\begin{equation}
P(Y_{B(\delta)}=0)=\delta+(1-\delta)(\alpha-\varepsilon)
\label{resulting YBdel}%
\end{equation}
Thus, $P(Y_{B(\delta)}=0)>\alpha$ iff $\delta(1-\alpha)>\varepsilon(1-\delta)$
which holds for $\varepsilon$ small enough. Thus
\begin{equation}
\max_{s\in\mathbb{S}}R(\delta,s)=1 \label{maxreg}%
\end{equation}
for all $\delta\in(0,1].$ If instead $\delta=0$ then nature chooses
$P(Y_{1}=1)=1$ and $P(Y_{0}=0)=1$ which leads to regret of 1.

This result may be surprising at first. For example, if the DM tries to be
completely balanced and picks $\delta=1/2$ (which seems reasonable in the no
data case) nature can inflict regret equal to one by picking $P(Y_{1}=0)=1$
and e.g., $P(Y_{0}=0)=\alpha-\min\{\alpha/2,(1-\alpha)/2\}.$ This is in stark
contrast to the case considered in Stoye (2009) where the DM cares about
expected outcome under $\delta,$ that is, $\mu_{0}(1-\delta)+\mu_{1}\delta,$
where $\mu_{t}$ for $t=1,2$ denotes the expectation of $Y_{t}$ under $s.$
Regret for given $\delta$ and $s$ (which only matters via $(\mu_{0},\mu_{1}))$
then equals $\max\{\mu_{0},\mu_{1}\}-(\mu_{0}(1-\delta)+\mu_{1}\delta).$ When
the DM picks a $\delta$ with $\delta\geq1/2$ then the maximal regret nature
can inflict equals $\delta$ obtained for any $s$ with $E_{s}Y_{0}=1$ and
$E_{s}Y_{1}=0.$ Therefore, the DM's minimax regret choice is $\delta=1/2.$
What explains the different results with quantiles and expectations? What
drives the results with quantiles is the discontinuity of the $\alpha
$-quantile of a random variable $X$ with respect to the cdf $F_{X}$ of $X.$ In
the above construction the random variables $Y_{0}$ and $Y_{B(\delta)}$ have
cdfs that are uniformly \textquotedblleft very close\textquotedblright\ yet
their $\alpha$-quantiles differ by 1. E.g. take again $\delta=1/2$ and
consider $\alpha=.99.$ Then both $Y_{0}$ and $Y_{B(\delta)}$ are Bernoulli
with $P(Y_{0}=0)=.99-.005=.985$ and $P(Y_{B(\delta)}=0)=1/2+1/2\cdot
.985=.9925.$ Thus, their cdfs are almost identical but their $.99$-quantiles
differ by 1. Instead the expectations of these two random variables are very
close.\medskip

Surprisingly, the intuition of the no-data example generalizes to cases with
arbitrary sample size $N>0.$ Again, for any given treatment rule $\delta
\in\mathbb{D}$ one constructs an $s_{\delta}\in\mathbb{S}$ for which
$\max\{q_{s_{\delta},\alpha}(Y_{0}),q_{s_{\delta},\alpha}(Y_{1})\}=1$ and
$u(\delta,s_{\delta})=0$ and thus $R(\delta,s_{\delta})=1.$ Again $s_{\delta
}\in\mathbb{S}$ can be chosen such that $Y_{0}$ and $Y_{1}$ are independent
and both have Bernoulli distributions. That is, $s_{\delta}$ is then fully
described by the two parameters $P_{s_{\delta}}(Y_{0}=0)$ and $P_{s_{\delta}%
}(Y_{1}=0).$ Denote by $0^{N_{0}}$ an $N_{0}$-dimensional column vector of zeros.

In case (i) of Proposition \ref{Analogue to Proposition 1 in Stoye 2009}, if
$\delta$ is such that $\delta((0^{N_{0}\prime},v^{\prime})^{\prime})=1$ for
all $v\in\{0,1\}^{N_{1}}$ then nature attempts punishing the DM for always
using treatment 1 when no successes are observed for treatment 0, by choosing
$P_{s_{\delta}}(Y_{1}=0)=1$ and by choosing $P(Y_{0}=0)$ \textquotedblleft big
enough\textquotedblright\ that only zeros are observed for the untreated
individuals \textquotedblleft sufficiently often\textquotedblright%
\ guaranteeing $Y_{B(\delta(w))}$ has $\alpha$-quantile 0, but small enough
that $Y_{0}$ has $\alpha$-quantile 1. The proof shows that this can indeed be
achieved by picking $P(Y_{0}=0)=\alpha-\varepsilon$ for some small enough
$\varepsilon>0.$

If on the other hand $\delta$ is such that $\delta((0^{N_{0}\prime},v^{\prime
})^{\prime})<1$ for at least one $v\in\{0,1\}^{N_{1}}$ then nature attempts
punishing the DM for not using treatment 1 often enough when no successes are
observed for treatment 0, by choosing $P_{s_{\delta}}(Y_{0}=0)=1$ and by
choosing $P(Y_{1}=0)$ small enough that $Y_{1}$ has $\alpha$-quantile 1 but
\textquotedblleft big enough\textquotedblright\ that $Y_{B(\delta(w))}$ has
$\alpha$-quantile 0. The proof shows that this can indeed be achieved by
picking $P(Y_{0}=0)=\alpha-\varepsilon$ for some small enough $\varepsilon>0.$

A surprising fact about the proof is that exploiting properties of $\delta$ on
the set of samples $\{(0^{N_{0}\prime},v^{\prime})^{\prime},$ $v\in
\{0,1\}^{N_{1}}\}$ alone gives nature enough leverage to inflict maximal
regret.\smallskip

3. In part (iii) of Proposition \ref{Analogue to Proposition 1 in Stoye 2009},
in the case $q_{\alpha}(Y_{0})<1/2$ we could not rule out that there are other
minimax regret rules besides $\delta^{1}\equiv1$ when $N>0.$ When $N=0,$ it is
obvious that $\delta^{1}$ is the unique minimax rule. A minimax regret
treatment rule, if it is not unique, may be inadmissible. In the case of
\textquotedblleft testing an innovation\textquotedblright\ $\delta^{0}\equiv0$
(when $q_{\alpha}(Y_{0})>1/2)$ is admissible, but we have not determined
whether that is true for $\delta^{1}$ (when $q_{\alpha}(Y_{0})<1/2)$ and for
$\delta^{.5}\equiv1/2$ (when $q_{\alpha}(Y_{0})=1/2).$\smallskip

4. In case $Y_{0},Y_{1}\in(0,1)$ rather than $Y_{0},Y_{1}\in\lbrack0,1],$
$\max_{s\in\mathbb{S}}R(\delta,s)$ does not generally exist when $\mathbb{S}$
denotes the set of all joint probability distribution over the potential
outcomes for $Y_{0},Y_{1}\in(0,1)$. E.g. in cases (i) and (ii) nature choosing
only distributions with support on $[\varepsilon,1-\varepsilon]$ for some
small $\varepsilon>0$ it can generate regret of $1-2\varepsilon$ and therefore
$\sup_{s\in\mathbb{S}}R(\delta,s)=1$ for the unrestricted space of
distributions for $Y_{0},Y_{1}\in(0,1).$ This result can be proven using the
exact same proof technique as for Proposition
\ref{Analogue to Proposition 1 in Stoye 2009}. Therefore, if minimax rules are
defined with $\sup_{s\in\mathbb{S}}R(\delta,s)$ (as we do in
(\ref{minimax rule})) rather than $\max_{s\in\mathbb{S}}R(\delta,s)$ the
results in Proposition \ref{Analogue to Proposition 1 in Stoye 2009}(i)-(ii)
continue to hold.\smallskip

5. If $Y_{0},Y_{1}\in\lbrack0,\infty)$ rather than $Y_{0},Y_{1}\in
\lbrack0,1],$ then $\max_{s\in\mathbb{S}}R(\delta,s)$ does not typically exist
when $\mathbb{S}$ denotes the set of all joint probability distributions over
the potential outcomes for $Y_{0},Y_{1}\in\lbrack0,\infty)$. E.g. in cases (i)
and (ii) nature choosing only distributions with support on $[0,M]$ for $M>0$
it can generate regret of $M$ and therefore $\sup_{s\in\mathbb{S}}%
R(\delta,s)=\infty$ when $\mathbb{S}$ denotes the set of all joint
distributions for $Y_{0},Y_{1}\in\lbrack0,\infty).$ This result can be proven
using the exact same proof technique as for Proposition
\ref{Analogue to Proposition 1 in Stoye 2009}. Therefore, if minimax rules are
defined with $\sup_{s\in\mathbb{S}}R(\delta,s)$ (as we do in
(\ref{minimax rule})) rather than $\max_{s\in\mathbb{S}}R(\delta,s)$ then for
$\alpha\in(0,1)$ any treatment rule is minimax regret optimal in cases (i)-(ii).

Given the limit experiment results in Hirano and Porter (2009) it would also
be interesting to study the case where potential outcomes are restricted to be
normally distributed. However, in that case, one would need to employ other
proof techniques than the ones used in the current paper.\medskip

\textbf{Restrictions on nature's action space}

In what follows we impose various restrictions on nature's action space and
study the implications on the results obtained in Proposition
\ref{Analogue to Proposition 1 in Stoye 2009}. For simplicity assume
$\alpha\in(0,1).$

\begin{corollary}
\label{Analogue to Proposition 1 in Stoye 2009 with restrictions}$($a$)$ The
results in Proposition \ref{Analogue to Proposition 1 in Stoye 2009}$($%
i$)$-$($ii$)$ remain valid when rather than $Y_{0},Y_{1}\in\lbrack0,1]$ the
setup is altered to $Y_{0},Y_{1}\in\{0,1\}.$ Similarly, the result in
$($iii$)$ remains valid if the distributions of $Y_{0}$ and $Y_{1}$ are
restricted to be discrete and supported on at most $N+1$ points in $[0,1]$.

$($b$)$ If $\mathbb{S}$ equals the set of distributions for $Y_{0},Y_{1}%
\in\lbrack0,1]$ whose marginals are all continuous $($with respect to Lebesgue
measure$)$ then Proposition \ref{Analogue to Proposition 1 in Stoye 2009}
continues to hold.
\end{corollary}

\textbf{Comments.} 1. Corollary
\ref{Analogue to Proposition 1 in Stoye 2009 with restrictions}(a) is a direct
corollary from the proof of Proposition
\ref{Analogue to Proposition 1 in Stoye 2009}. In the proof of Proposition
\ref{Analogue to Proposition 1 in Stoye 2009}(i)-(ii) only Bernoulli
distributions are used for nature while in part (iii) only discrete
distributions are used.

2. Corollary \ref{Analogue to Proposition 1 in Stoye 2009 with restrictions}%
(b) considers the case where nature is restricted to continuous distributions.
We have seen in Proposition \ref{Analogue to Proposition 1 in Stoye 2009} that
in cases (i)-(ii) any treatment rule $\delta\in\mathbb{D}$ is minimax and
$\max_{s\in\mathbb{S}}R(\delta,s)=1.$ Because without pointmasses it is
impossible for a random variable to have $\alpha$-quantile equal to 0 it
follows that $R(\delta,s)$ is always strictly smaller than 1 for any pair
$(\delta,s)$ when $s$ is continuous with respect to Lebesgue measure. The main
construction in the proof of Proposition
\ref{Analogue to Proposition 1 in Stoye 2009} still goes through when one
considers a sequence of continuous distributions that converge to the
Bernoulli distributions that are used in that proof. As a technical detail it
is important to use $\sup_{s\in\mathbb{S}}R(\delta,s)$ rather than $\max
_{s\in\mathbb{S}}R(\delta,s)$ in the definition of regret, because $\max
_{s\in\mathbb{S}}R(\delta,s)$ would not exist in the case considered here. One
can establish that $\sup_{s\in\mathbb{S}}R(\delta,s)=1$ for any treatment rule
$\delta\in\mathbb{D}\mathfrak{.}$

3. In the case of \emph{\textquotedblleft testing an
innovation\textquotedblright} a restriction on nature's action space occurs if
one assumes that the entire distribution of $Y_{0}$ is known, not just its
$\alpha$-quantile. Namely, in that case the set $\mathbb{S}$ consists of all
possible distributions $s$ for $Y_{1}\in\lbrack0,1]$ (while the distribution
of $Y_{0}$ is given)$.$ The analysis of that problem is more difficult. Denote
by $F_{Y_{0}}$ and $q_{\alpha}(Y_{0})$ the cdf and $\alpha$-quantile of
$Y_{0},$ respectively.

Take $r=0$ in (\ref{definition of the alpha quantile}) and assume $F_{Y_{0}%
}(q_{\alpha}(Y_{0}))>\alpha$. Consider the case $N=0$ in which case a
treatment rule $\delta\in\lbrack0,1]$ denotes the treatment probability. Under
these assumptions the following statements hold.

Denote by $q(\delta)$ the smallest $q\in\lbrack0,q_{\alpha}(Y_{0})]$ such
that
\begin{equation}
\delta+(1-\delta)F_{Y_{0}}(q)\geq\alpha. \label{condition}%
\end{equation}
Then, when $q_{\alpha}(Y_{0})<1/2,$ $\delta^{1}\equiv1$ is the only minimax
regret rule and $\max_{s\in\mathbb{S}}R(\delta^{1},s)=q_{\alpha}(Y_{0}).$ When
$q_{\alpha}(Y_{0})=1/2$ any $\delta\in\lbrack0,1]$ is minimax regret with
maximal regret equal to $q_{\alpha}(Y_{0}).$ Finally, when $q_{\alpha}%
(Y_{0})>1/2,$ $\delta\in\lbrack0,1]$ is minimax regret iff
\begin{equation}
q_{\alpha}(Y_{0})-q(\delta)\leq1-q_{\alpha}(Y_{0}).
\label{optimality condition}%
\end{equation}
Given that $q_{\alpha}(Y_{0})-q(\delta)$ is a weakly increasing function in
$\delta,$ if we denote by $\delta^{\ast}\in\lbrack0,1]$ an intersection of
$q_{\alpha}(Y_{0})-q(\delta)$ and $1-q_{\alpha}(Y_{0})$ then any $\delta
\in\lbrack0,\delta^{\ast}]$ is minimax regret and for those $\delta$ we have
$\max_{s\in\mathbb{S}}R(\delta,s)=1-q_{\alpha}(Y_{0})$.

In the Appendix, we give a proof of the statements made above. The results for
the case $q_{\alpha}(Y_{0})>1/2$ are partly in contrast to Proposition
\ref{Analogue to Proposition 1 in Stoye 2009}(iii) where there is a
\emph{unique} minimax regret rule. However, maximal regret is the same here as
in Proposition \ref{Analogue to Proposition 1 in Stoye 2009}(iii). We have not
yet generalized the results to arbitrary sample sizes $N>0$ and to the case
$F_{Y_{0}}(q_{\alpha}(Y_{0}))=\alpha.$

\subsection{Finite sample simulation\label{finite sample simulation}}

For the case of \textquotedblleft testing an innovation\textquotedblright%
\ Proposition \ref{Analogue to Proposition 1 in Stoye 2009}(iii) establishes
that the minimax regret criterion when applied to $\alpha$-quantiles does not
favor data-driven rules. In this section we conduct a simulation experiment to
juxtapose the pointwise (in $s$) regret of the data-driven empirical success
rule $\delta^{ES},$ defined by
\begin{equation}
\delta^{ES}(w)=I(q_{\alpha}(Y_{1},w)>q_{\alpha}(Y_{0}))+.5I(q_{\alpha}%
(Y_{1},w)=q_{\alpha}(Y_{0})), \label{ESR}%
\end{equation}
where $q_{\alpha}(Y_{1},w)$ denotes the $\alpha$-sample quantile of $Y_{1}$
for the sample $w,$ with the regret of the minimax regret rule $\delta
^{1}\equiv1$ (in the case when $q_{\alpha}(Y_{0})\leq1/2),$ $\delta^{.5}%
\equiv1/2$ (in the case when $q_{\alpha}(Y_{0})=1/2),$ and the minimax regret
rule $\delta^{0}\equiv0$ (in the case when $q_{\alpha}(Y_{0})\geq1/2)$.

Obviously, when simulating regret we cannot possibly include all states of
nature $s\in\mathbb{S}$. For the simulation experiment, we create a
\textquotedblleft sufficiently rich\textquotedblright\ subset of distributions
for $(Y_{0},Y_{1}).$ Namely, for some $n,w\in\mathbb{N}$, we consider the set
of states of nature $\mathbb{S}^{E}=\mathbb{S}^{E}(n,w)$ that consists of all
discrete distributions $s$ for $Y_{1}$ supported on a grid
\begin{equation}
\{0,1/n,2/n,...,n/n\} \label{grid}%
\end{equation}
with probabilities
\begin{equation}
P_{s}(Y_{1}=j/n) \label{probs}%
\end{equation}
being of the form $i/w,$ $i=0,...,w;$ while for the distribution of $Y_{0}$ we
consider two choices, namely

I) $Y_{0}\in\{0,q_{\alpha}(Y_{0})\}$ with $P(Y_{0}=0)=\alpha-\varepsilon$ and
$P(Y_{0}=q_{\alpha}(Y_{0}))=1-(\alpha-\varepsilon)$ for $\varepsilon=.000001$ and

II) $Y_{0}$ being continuously distributed on [0,1] with density $f(x)$ equal
to $\alpha/q_{\alpha}(Y_{0})$ for $x\leq q_{\alpha}(Y_{0})$ and equal to
$(1-\alpha)/(1-q_{\alpha}(Y_{0}))$ otherwise. Note that in both I) and II) the
$\alpha$-quantile of $Y_{0},$ $q_{s,\alpha}(Y_{0}),$ equals $q_{\alpha}%
(Y_{0}).$ We report results for all choices of $\alpha\in\{.1,.5,.9\}$,
$q_{\alpha}(Y_{0})\in\{.1,.5,.9\},$ sample size $N=30,$ and $(n,w)=(6,12).$
The latter leads to $S^{E}$ having cardinality 18564.

For the given choices of $n,w,$ and $N$ and for each choice of $\alpha$ and
$q_{\alpha}(Y_{0})$, for each state of nature $s\in\mathbb{S}^{E}(n,w)$ and
one (of the two possible) choice of distribution for $Y_{0}$ we simulate
regret for the four treatment rules $\delta^{ES},$ $\delta^{1},$ $\delta
^{.5},$ and $\delta^{0}$ by generating $R=100K$ samples of size $N$ by drawing
i.i.d. observations of the distribution of $Y_{1}.$ We use $r=0$ when
simulating $\alpha$-quantiles of the outcome distribution under the various
treatment rules. We analytically calculate $\alpha$-quantiles for $Y_{1}$ for
$Y_{B(\delta^{.5})}$, and likewise use the true $\alpha$-quantile $q_{\alpha
}(Y_{0})$ of $Y_{0}$ when calculating regret. For a given treatment rule
$\delta$ and a given state of nature $s$, regret is calculated as
$R(\delta,s)=\max\{q_{\alpha}(Y_{0}),q_{s,\alpha}(Y_{1})\}-q_{s,\alpha
}(Y_{B(\delta(w))}).$

We compare the treatment rules along several dimensions, namely

a) mean regret over all 18564 states of nature $s\in\mathbb{S}^{E}(n,w)$,

b) maximal regret over all states of nature $s\in\mathbb{S}^{E}(n,w)$,

c) minimal regret over all states of nature $s\in\mathbb{S}^{E}(n,w)$, and

d) the proportion of $s\in\mathbb{S}^{E}(n,w)$ for which regret for the
empirical success rule $\delta^{ES}$ is smaller than regret for each one of
its three competitors.

Just for clarity, in a) for each treatment rule we sum up its regret over all
18564 states of nature $s\in\mathbb{S}^{E}(n,w)$ and then report that sum
divided by 18564. For a given $s\in\mathbb{S}^{E}(n,w)$, in our simulations,
we interpret regret of $\delta^{ES}$ as smaller than regret of another rule,
say $\delta^{0},$ if the simulated regret of $\delta^{ES}$ is smaller than the
simulated regret of the rule $\delta^{0}$ minus a threshold of $\xi.$ If
instead the simulated regret of $\delta^{ES}$ falls into the interval
[(simulated regret of $\delta^{0})-\xi,$(simulated regret of $\delta^{0})+\xi
$] we record regrets of the two rules as equal for that state of nature. We
take $\xi=.00000001$ below. Similarly, when programming the empirical success
rule in (\ref{ESR}) the event \textquotedblleft$q_{\alpha}(Y_{1},w)=q_{\alpha
}(Y_{0})\textquotedblright$ is implemented as the simulated $\alpha$-quantile
of $Y_{1}$ falling into the interval $[q_{\alpha}(Y_{0})-\xi,q_{\alpha}%
(Y_{0})+\xi].$

All reported results are rounded to the second digit after the comma and so a
reported zero could be as large as .004.

In the tables below we do not include results for c) because those results
turn out to be equal to zero for all treatment rules and all designs except
for $\delta^{.5}$ when $q_{\alpha}(Y_{0})\in\{.1,.9\}$ in which case minimal
regret equals .067 for all choices of $\alpha$ and both choices of
distributions for $Y_{0}.$

TABLE I provides results for maximal and mean regret over all 18564 states of
nature $s\in S^{E}(6,12)$ for the four different treatment rules. Results for
rules that are minimax regret in a given setting are reported in bold.

We first discuss results for maximal regret. The treatment rules that are
known to be minimax regret for unrestricted $\mathbb{S}$ also have smallest
maximal regret over $\mathbb{S}^{E}(n,w)$ (relative to the other treatment
rules considered here) for all cases except for case II) with $\alpha=.9,$
when $q_{\alpha}(Y_{0})=.1$ (in which case $\delta^{ES}$ has smaller maximal
regret than the optimal rule $\delta^{1})$, when $q_{\alpha}(Y_{0})=.5$ (in
which case it is known that all four rules are optimal but in finite samples
for $\mathbb{S}^{E}(n,w)$ again $\delta^{ES}$ does best), and finally when
$q_{\alpha}(Y_{0})=.1$ (in which case $\delta^{ES}$ has smaller maximal regret
than the optimal rule $\delta^{0})$. The explanation is of course that
$\mathbb{S}^{E}(n,w)$ does not contain those states of nature that would
inflict the highest regret on $\delta^{ES}$ (like the particular Bernoulli and
discrete distributions for $Y_{0}$ and $Y_{1},$ respectively, that are used in
the proof of Proposition \ref{Analogue to Proposition 1 in Stoye 2009}(iii)).
Furthermore, reported finite sample maximal regret over $\mathbb{S}^{E}(n,w)$
for the optimal rules matches the theoretical value $\min\{q_{\alpha}%
(Y_{0}),1-q_{\alpha}(Y_{0})\}$ reported in Proposition
\ref{Analogue to Proposition 1 in Stoye 2009}(iii) except for the case II)
with $\alpha=.9$ when $q_{\alpha}(Y_{0})=.5$ where the optimal rule
$\delta^{ES}$ has maximal regret smaller than .5 (which occurs, again, because
$\mathbb{S}^{E}(n,w)$ is not rich enough). An open question from Proposition
\ref{Analogue to Proposition 1 in Stoye 2009}(iii) is whether for $q_{\alpha
}(Y_{0})<.5$ other minimax regret rules besides $\delta^{1}$ might exist. The
simulations for $q_{\alpha}(Y_{0})=.1$ are compatible with the possibility
that when $\alpha=.5$ or $.9$ also $\delta^{ES}$ might be minimax regret.

We next discuss results for mean regret. In most cases where $q_{\alpha}%
(Y_{0})\in\{.1,.9\}$ in which case either $\delta^{0}$ or $\delta^{1}$ are
minimax regret, their mean performance is also best (or very close to best)
among the four treatment rules. On the other hand, when $q_{\alpha}(Y_{0})=.5$
(in which case all four treatment rules are minimax regret according to
Proposition \ref{Analogue to Proposition 1 in Stoye 2009}(iii)) we see huge
difference in mean performance across the four treatment rules for a given
case I) or II) and $\alpha,$ but also huge differences in performance for a
given treatment rule and $\alpha$ across cases I) and II). With respect to the
former point, in case I) when $\alpha=.1$ mean regret for the four rules are
in the interval [0,.43]. With regards to the latter point, e.g., for
$\delta^{ES}$ when $\alpha=.1$ mean regret equals .43 and .06, in cases I) and
II) respectively. Differences across cases I) and II) are also often huge for
other quantiles. E.g. again for $\delta^{ES}$ when $q_{\alpha}(Y_{0})=.9$ and
$\alpha=.1$ mean regret equals .9 and 0, in cases I) and II) respectively.
(Recall that we round results to the second digit. When $q_{\alpha}(Y_{0})=.9$
and $\alpha=.1$ there are not many distributions for $Y_{1}$ in $\mathbb{S}%
^{E}(n,w)$ that have an $\alpha$-quantile that exceeds .9.)\smallskip

We next discuss the results for exercise d) contained in TABLE II where we
report the proportion of the 18564 states of nature $s\in S^{E}(6,12)$ for
which the regret of $\delta^{ES}$ is smaller than the regret of $\delta$ where
consider all $\delta$ from the set of no-data rules $\{\delta^{1},\delta
^{.5},\delta^{0}\}$. The results indicate that for many states $s$, choices of
$q_{\alpha}(Y_{0})$, $\alpha,$ and the distribution for $Y_{0},$
$R(\delta^{ES},s)$ and $R(\delta,s)$ are very close which leads to numerically
unstable results. To deal with the instability we introduce the buffer $\xi$
as explained above and $Prop(R(\delta^{ES},s)<R(\delta,s))$ and $Prop(R(\delta
^{ES},s)\leq R(\delta,s))$ in TABLE II\ represent the proportion of states $s$
for which $R(\delta^{ES},s)<R(\delta,s)-\xi$ and $R(\delta^{ES},s)<R(\delta
,s)+\xi,$ respectively. Those two proportions can be drastically different,
suggesting that in many scenarios regret for $\delta^{ES}$ and $\delta$ are
(virtually) identical. According to the measure $Prop(R(\delta^{ES},s)\leq
R(\delta,s)),$ maybe somewhat surprisingly given the results from TABLE I,
$\delta^{ES}$ is to be preferred over the other three rules in all scenarios
considered in Case II) (except the case $\alpha=q_{\alpha}(Y_{0})=.9$ for
$\delta^{0})$ and in the majority of scenarios in Case I) (except compared to
$\delta^{0}$ when $q_{\alpha}(Y_{0})=.5,\alpha=.1$ and except for most cases
with $q_{\alpha}(Y_{0})=.9$ and all three rules$).$ Quite often $Prop(R(\delta
^{ES},s)\leq R(\delta,s))$ is reported as higher than 95\%, but the
improvement in regret for many states of nature is minuscule. For example
compared to $\delta^{.5}$ in Case I) with $\alpha=q_{\alpha}(Y_{0})=.1$ only
in 33.3\% of the cases $Prop(R(\delta^{ES},s)<R(\delta,s))$ while for 100\% of
the cases $Prop(R(\delta^{ES},s)\leq R(\delta,s))$ implying that the regret of
$\delta^{ES}$ in 66.4\% of the cases is at most $\xi$ smaller than the regret
of $\delta^{.5}$.

\section{Treatment choice with a covariate\label{covariate included}}

Next, as in Stoye (2009) we next allow for a discrete covariate $X\in
\mathcal{X}=\{x_{1},...,x_{K}\}$ that is observed both in the sample and in
the treatment data. Outcomes $Y_{t,x}$ now carry a double subindex to indicate
the treatment status $t\in\{0,1\}$ and the value of the covariate
$x\in\mathcal{X}.$ It is assumed that $x_{k}$ occurs with positive probability
for each $k=1,...,K.$ Denote by $F_{X}$ the distribution of $X.$ A state of
the world $s\in\mathbb{S}$ now represents a joint distribution for
$(Y_{t,x})_{t\in\{0,1\},\text{ }x\in\mathcal{X}}.$

A sample $w=w_{N}$ now consists of realizations $(t_{i},x_{i},y_{i})$ for
$i=1,...,N$ of $(T,X,Y_{T,X}),$ where $y_{i}$ is a realization of
$Y_{t_{i},x_{i}}$ and we consider again sampling designs (i)-(iii) from
Section \ref{No covariates}. It is assumed that $x_{i},$ $i=1,...,n,$ are
i.i.d. and $F_{X}$ is independent of $s$ and $T$.\smallskip

In design (i), \textquotedblleft fixed number of treated/untreated
units\textquotedblright, there are $N_{0}$ observations $(0,x_{i},y_{i})$ with
$y_{i}$ being i.i.d. draws of $Y_{0,x_{i}},$ $i=1,...,N_{0}$ and
$N_{1}=N-N_{0}$ observations $(1,x_{i},y_{i})$ with $y_{i}$ being i.i.d. draws
of $Y_{1,x_{i}},$ $i=N_{0}+1,...,N.$ Note that there are not typically equally
many treated and untreated observations for each covariate $x_{k},$
$k=1,...,K.$ In fact, there may be zero observations altogether in the sample
for a given $x_{k}.$

In design (ii), \textquotedblleft random sampling\textquotedblright, the
observations $(t_{i},x_{i},y_{i})$ for $i=1,...,N$ are i.i.d. realizations of
$(T,X,Y_{T,X})$ with $P(T_{i}=1)=p\in(0,1)$ with $T,X$, and $s$ being
independent of each other$.$

In design (iii), \textquotedblleft testing an innovation\textquotedblright,
$(1,x_{i},y_{i})$ for $i=1,...,N$ are observed where $y_{i}$ is an independent
realization of $Y_{1,x_{i}},$ for $i=1,...,N.$

In design (iii), we assume the DM knows the $\alpha$-quantile, denoted by
$q_{\alpha}(Y_{0,X}),$ of $Y_{0,X}$ and nature can choose from joint
distributions $s\in\mathbb{S}$ for $(Y_{t,x})_{t\in\{0,1\},\text{ }%
x\in\mathcal{X}}$ such that the $\alpha$-quantile of $Y_{0,X}$ equals
$q_{\alpha}(Y_{0,X}).$\smallskip

A treatment rule $\delta$ maps a sample $w_{N}$ onto treatment probabilities
for each $x_{k}$ for $k=1,...,K,$ that is $\delta(w_{N})\in\lbrack0,1]^{K},$
where the $k$-th component of $\delta(w_{N})$ indicates the treatment
probability for individuals with covariate $x_{k}$ for $k=1,...,K.$ Denote the
$k$-th component of $\delta(w_{N})$ by $\delta_{x_{k}}(w_{N})$ for
$k=1,...,K.$ With some abuse of notation, for each design (i)-(iii) we denote
the set of all treatment rules by the same symbols $\mathbb{D}$ (even though
it means something different for different designs)$.$

The object of interest is
\begin{equation}
u(\delta,s)=q_{s,F_{X},\alpha}(Y_{B(\delta_{X}(w_{N})),X})
\label{obj of interest with cov}%
\end{equation}
the $\alpha$-quantile of the outcome distribution. With that definition,
regret is then defined formally analogously to the setup without a covariate,
namely, $R(\delta,s)=\sup_{d\in\mathbb{D}}u(d,s)-u(\delta,s).$ Alternatively,
one could focus on the $\alpha$-quantile of the outcome distribution for a
particular covariate only, $x_{1}$ say, $u_{x_{1}}(\delta,s)=q_{s,F_{X}%
,\alpha}(Y_{B(\delta_{x_{1}}(w_{N})),x_{1}})$ and obtain analogous results to
the ones in Corollary \ref{statement with covariates} below.

If the space of probability distributions for nature $\mathbb{S}$ is
unrestricted and thus equals the space of all distributions for $(Y_{t,x}%
)_{t\in\{0,1\},\text{ }x\in\mathcal{X}}$ one can show in designs (i)-(ii) (as
an implication of the proof of Proposition
\ref{Analogue to Proposition 1 in Stoye 2009}) that for every treatment rule
$\delta$ maximal risk over $\mathbb{S}$ equals 1, $\max_{s\in\mathbb{S}%
}R(\delta,s)=1$. This result then immediately implies that every treatment
rule is minimax regret. The corollary (to Proposition
\ref{Analogue to Proposition 1 in Stoye 2009}) that follows gives a stronger
result; it shows that maximal risk continues to be 1 even if certain
restrictions are imposed on $\mathbb{S}$. For simplicity of the presentation
assume $\alpha\in(0,1).$

\begin{corollary}
\label{statement with covariates}$($i$)$-$($ii$)$ In the case of the designs
\textquotedblleft fixed number of treated/untreated units\textquotedblright%
\ and \textquotedblleft random sampling\textquotedblright\ for any treatment
rule $\delta\in\mathbb{D}$, $\max_{s\in\mathbb{S}}R(\delta,s)=1$ if
$\mathbb{S}$ includes as a subset all joint distributions for $(Y_{t,x}%
)_{t\in\{0,1\},\text{ }x\in\mathcal{X}}$ whose marginals are independent
Bernoulli distributions. Therefore, \emph{any} $\delta\in\mathbb{D}$ is
minimax regret.

$($iii$)$ In the case of \textquotedblleft testing an
innovation\textquotedblright\ assume the DM knows the $\alpha$-quantile,
denoted by $q_{\alpha}(Y_{0,X}),$ of $Y_{0,X}.$ If $q_{\alpha}(Y_{0,X})=1/2$
then any $\delta\in\mathbb{D}$ is minimax regret; if $q_{\alpha}(Y_{0,X})>1/2$
then $\delta^{0}=0$ is the unique minimax regret rule, and if $q_{\alpha
}(Y_{0,X})<1/2$ then $\delta^{1}=1$ is a minimax regret rule. In each case,
$max_{s\in\mathbb{S}}R(\delta,s)=\min\{q_{\alpha}(Y_{0,X}),1-q_{\alpha
}(Y_{0,X})\}$.
\end{corollary}

\textbf{Comments:} 1. Various variants could be considered in design (iii).
E.g. one could instead assume that the joint distribution $(Y_{0,x}%
)_{x\in\mathcal{X}}$ is known (in which case nature only chooses a joint
distribution for $(Y_{1,x})_{x\in\mathcal{X}})$ or one could assume that the
DM knows the vector $(q_{\alpha}(Y_{0,x_{1}}),...,q_{\alpha}(Y_{0,x_{K}}))$ of
$\alpha$-quantiles of all the marginal distributions $(Y_{0,x})_{x\in
\mathcal{X}}$ and nature can choose from joint distributions $s\in\mathbb{S}$
for $(Y_{t,x})_{t\in\{0,1\},\text{ }x\in\mathcal{X}}$ such that all the
marginals $(Y_{0,x})_{x\in\mathcal{X}}$ have the required $\alpha$-quantiles.

2. One could also consider alternative sampling designs (i)-(iii), where in
the sampling stage, rather than being randomly assigned, the values of the
covariate $X$ are assigned deterministically, as is done for treatment status
in design (i). This could be referred to as \textquotedblleft stratified
sampling.\textquotedblright

For brevity we do not explicitly deal with these variations.

\section{\textbf{Conclusion}}

We derive minimax regret treatment rules in finite samples when an $\alpha
$-quantile of the outcome distribution is the focus of interest. We establish
that when the sample i) consists of a fixed number of untreated/treated units
or ii) is generated via random treatment assignment then \emph{all} treatment
rules are minimax regret and therefore the minimax regret criterion is not
helpful in singling out a recommended treatment rule. Given that the same
shortcoming applies to the max-min criterion, an important question concerns
finding a meaningful criterion in this setup based on which an optimal
treatment rule should be chosen. The idea from Montiel Olea, Qiu, and Stoye
(2023) to look for rules that randomize \textquotedblleft the
least\textquotedblright\ in a situation where there are multiple minimax
regret rules would not lead to a unique rule in our setup because in cases i)
and ii)\ both $\delta^{0}$ and $\delta^{1}$ are minimax regret and never
randomize. Given all these facts, it then seems reasonable to simply adopt a
rule that is minimax regret optimal when regret is based on the notion of
\emph{expected} welfare (in particular, such a rule is optimal according to
criterion (\ref{minimax rule})).

We also establish that when iii) the sample consists of only realizations from
the treated population while the $\alpha$-quantile of the untreated population
is known, never treating is the unique minimax rule if the known quantile
exceeds .5, while always treating is a minimax rule if that quantile is
strictly smaller than .5, and finally, if the quantile equals .5, then any
treatment rule is minimax regret. It follows that based on the minimax regret
criterion, rules that take the data into consideration are never strictly preferred.

An interesting but quite difficult extension that we are currently
investigating concerns applying the minimax regret criterion in finite samples
to conditional value at risk, defined as $S_{s,\alpha}(Y):=\alpha^{-1}%
E_{s}[Y1(Y\leq q_{s,\alpha}(Y))]$ for a random variable $Y$ rather than to the
$\alpha$-quantile of $Y$.\footnote{Or, using the more general definition
$S_{s,\alpha}(Y):=\sup_{\gamma\in R}\{\gamma-\alpha^{-1}E_{s}[\gamma-Y]_{+}%
\}$, see Qi, Pang, and Liu (2023) were, $[Y]_{+}=\max\{Y,0\}$ denotes the
positive part of $Y$.}

\section{Appendix}

\textbf{Proof of Proposition \ref{Analogue to Proposition 1 in Stoye 2009}.
}We use the notation $P_{s}(A)$ to denote probability of the event $A$ when
nature chooses the state of the world $s\in\mathbb{S}$ and we write shorthand
$w$ instead of $w_{N}$ for the sample of size $N$. Note that for a state of
the world $s\in\mathbb{S}$ and treatment rule $\delta\in\mathbb{D}$%
\begin{align}
&  P_{s}(Y_{B(\delta(w))}\overset{}{\leq}q)\nonumber\\
&  \overset{}{=}P_{s}(B(\delta(w))\overset{}{=}1\text{ }\&\text{ }%
Y_{1}\overset{}{\leq}q)+P_{s}(B(\delta(w))\overset{}{=}0\text{ }\&\text{
}Y_{0}\overset{}{\leq}q)\nonumber\\
&  \overset{}{=}P_{s}(B(\delta(w))\overset{}{=}1)P_{s}(Y_{1}\overset{}{\leq
}q)+P_{s}(B(\delta(w))\overset{}{=}0)P_{s}(Y_{0}\overset{}{\leq}q)
\label{prob derivation}%
\end{align}
and analogously for $P_{s}(Y_{B(\delta(w))}\geq q)$, where the second equality
uses independence. By definition an $\alpha$-quantile $q$ of $Y_{B(\delta
(w))}$ satisfies $P_{s}(Y_{B(\delta(w))}\leq q)\geq\alpha$ and $P_{s}%
(Y_{B(\delta(w))}\geq q)\geq1-\alpha.$\medskip

The proofs of (i)-(ii)\ proceed by showing that for any treatment rule
$\delta\in\mathbb{D}$ there exists a state of the world $s_{\delta}%
\in\mathbb{S}$ for which $R(\delta,s_{\delta})=1.$ Because it is also true
that for any treatment rule $\delta\in\mathbb{D},$ $\max_{s\in\mathbb{S}%
}R(\delta,s)\leq1$ it follows that any treatment rule is minimax.

Lemma \ref{max utility}(ii) below establishes the (unsurprising) result that
for any arbitrary $s\in\mathbb{S}$
\begin{equation}
\max_{d\in\mathbb{D}}u(d,s)=\max_{d\in\mathbb{D}}q_{s,\alpha}(Y_{B(d(w))}%
)=\max\{q_{s,\alpha}(Y_{0}),q_{s,\alpha}(Y_{1})\}. \label{max quantile claim}%
\end{equation}
This result implies a simplified formula for $R(\delta,s)$ that we will use
from now on.\smallskip

In the proof of (i)-(ii) that follows, for any given treatment rule $\delta
\in\mathbb{D}$ we will construct a $s_{\delta}\in\mathbb{S}$ for which
$\max\{q_{s_{\delta},\alpha}(Y_{0}),q_{s_{\delta},\alpha}(Y_{1})\}=1$ and
$u(\delta,s_{\delta})=0.$ Namely, throughout the proof of (i)-(ii) for an
arbitrary treatment rule $\delta\in\mathbb{D}$ we construct a $s_{\delta}%
\in\mathbb{S}$ whose independent marginals for $Y_{0}$ and $Y_{1}$ are
Bernoulli (supported on $\{0,1\}$) and for which $R(\delta,s_{\delta})=1$.
Under these restrictions, the distribution $s_{\delta}$ for $(Y_{0},Y_{1})$ is
then fully defined by
\begin{equation}
a_{0}:=P_{s_{\delta}}(Y_{0}=0)\text{ and }a_{1}:=P_{s_{\delta}}(Y_{1}=0).
\label{definition of s}%
\end{equation}
The $\alpha$-quantiles of $Y_{0},$ $Y_{1},$ and $Y_{B(\delta(w))}$ in the
constructions below will be unique when $\alpha\in(0,1)$. Therefore, the
particular value of $r\in\lbrack0,1]$ has no impact on the results.

\textbf{Part (i).} As in (\ref{prob derivation}) and with $s_{\delta}$ just
defined%
\begin{align}
&  P_{s_{\delta}}(Y_{B(\delta(w))}\overset{}{=}0)\nonumber\\
&  =%
{\textstyle\sum\nolimits_{u\in\{0,1\}^{N_{0}}}}
{\textstyle\sum\nolimits_{v\in\{0,1\}^{N_{1}}}}
a_{0}^{N_{0}-\overline{u}}(1-a_{0})^{\overline{u}}a_{1}^{N_{1}-\overline{v}%
}(1-a_{1})^{\overline{v}}[\delta((u^{\prime},v^{\prime})^{\prime}%
)a_{1}+(1-\delta((u^{\prime},v^{\prime})^{\prime}))a_{0}]\nonumber\\
&  =%
{\textstyle\sum\nolimits_{i=0}^{N_{0}}}
{\textstyle\sum\nolimits_{j=0}^{N_{1}}}
a_{0}^{N_{0}-i}(1-a_{0})^{i}a_{1}^{N_{1}-j}(1-a_{1})^{j}[\delta^{i,j}%
a_{1}+(\binom{N_{0}}{i}\binom{N_{1}}{j}-\delta^{i,j})a_{0}],
\label{derivation}%
\end{align}
where for a vector $u\in\{0,1\}^{N_{0}}$ we denote by $\overline{u}%
\in\mathbb{N}$ the sum of its components (that is the number of 1's in the
vector) and similar for other vectors, and
\begin{equation}
\delta^{i,j}:=%
{\textstyle\sum\nolimits_{\substack{u\in\{0,1\}^{N_{0}},v\in\{0,1\}^{N_{1}%
}\\\overline{u}=i\text{ and }\overline{v}=j}}}
\delta((u^{\prime},v^{\prime})^{\prime}). \label{shorthand}%
\end{equation}
Note that in the second line of (\ref{derivation}), the term $a_{0}%
^{N_{0}-\overline{u}}(1-a_{0})^{\overline{u}}a_{1}^{N_{1}-\overline{v}%
}(1-a_{1})^{\overline{v}}$ is the probability to observe particular vectors
$u$ and $v$ while the term $\delta((u^{\prime},v^{\prime})^{\prime}%
)a_{1}+(1-\delta((u^{\prime},v^{\prime})^{\prime}))a_{0}$ is the probability
that an outcome 0 is reached when the sample consists of $u$ and
$v.$\smallskip

Assume $\alpha\in(0,1)$ first. When in the definition of $s_{\delta}$ in
(\ref{definition of s}) we take $(a_{0},a_{1})=(1,\alpha)$ we obtain
\begin{align}
P_{s_{\delta}}(Y_{B(\delta(w))}\overset{}{=}0)  &  =%
{\textstyle\sum\nolimits_{j=0}^{N_{1}}}
\alpha^{N_{1}-j}(1-\alpha)^{j}[\delta^{0,j}\alpha+(\binom{N_{1}}{j}%
-\delta^{0,j}))]\nonumber\\
&  =%
{\textstyle\sum\nolimits_{j=0}^{N_{1}}}
[\alpha^{N_{1}-j}(1-\alpha)^{j}(\alpha-1)]\delta^{0,j}+%
{\textstyle\sum\nolimits_{j=0}^{N_{1}}}
\alpha^{N_{1}-j}(1-\alpha)^{j}\binom{N_{1}}{j}. \label{special case}%
\end{align}
Given that the coefficient $\alpha^{N_{1}-j}(1-\alpha)^{j}(\alpha-1)$ in front
of $\delta^{0,j}$ is negative (noting that $\alpha\in(0,1))$ it follows that
$P_{s_{\delta}}(Y_{B(\delta(w))}=0)$ is strictly decreasing in each
$\delta^{0,j}.$ Therefore, we obtain
\begin{equation}
P_{s_{\delta}}(Y_{B(\delta(w))}=0)\geq%
{\textstyle\sum\nolimits_{j=0}^{N_{1}}}
\alpha^{N_{1}-j}(1-\alpha)^{j}\alpha\binom{N_{1}}{j}=\alpha,
\label{case1alpha2}%
\end{equation}
where the right hand bound follows from (\ref{special case}) by replacing
$\delta^{0,j}$ by its maximal value $\binom{N_{1}}{j}.$

Consider first the case where at least one of the $\delta^{0,j}$ (for
$j=0,...,N_{1}$) is smaller than $\binom{N_{1}}{j}.$ In that case it follows
that $P_{s_{\delta}}(Y_{B(\delta(w))}=0)>\alpha.$ Furthermore, notice that
$P_{s_{\delta}}(Y_{B(\delta(w))}=0)$ is continuous in $a_{1}$ and therefore
$P_{s_{\delta}}(Y_{B(\delta(w))}=0)>\alpha$ when instead of $(a_{0}%
,a_{1})=(1,\alpha),$ $s_{\delta}$ is taken as $(a_{0},a_{1})=(1,\alpha
-\varepsilon)$ for some small enough $\varepsilon>0.$ But then for that choice
of $s_{\delta},$ $u(\delta,s_{\delta})=0$ and $\max\{q_{s_{\delta},\alpha
}(Y_{0}),q_{s_{\delta},\alpha}(Y_{1})\}=q_{s_{\delta},\alpha}(Y_{1})=1.$
Therefore $R(\delta,s_{\delta})=1$ as desired.

Next, consider the case where $\delta^{0,j}=\binom{N_{1}}{j}$ for all
$j=0,...,N_{1}$ (which implies that $\delta((0^{N_{0}\prime},v^{\prime
})^{\prime})=1$ for all $v\in\{0,1\}^{N_{1}},$ where $0^{N_{0}}$ denotes an
$N_{0}$-dimensional column vector of zeros).\ In that case, from
(\ref{derivation}) we obtain with $(a_{0},a_{1})=(\alpha,1)$
\begin{align}
&  P_{s_{\delta}}(Y_{B(\delta(w))}\overset{}{=}0)\nonumber\\
&  \overset{}{=}%
{\textstyle\sum\nolimits_{i=0}^{N_{0}}}
\alpha^{N_{0}-i}(1-\alpha)^{i}[\delta^{i,0}+(\binom{N_{0}}{i}-\delta
^{i,0}))\alpha]\nonumber\\
&  \overset{}{=}\alpha^{N_{0}}+%
{\textstyle\sum\nolimits_{i=1}^{N_{0}}}
\alpha^{N_{0}-i}(1-\alpha)^{i}[\delta^{i,0}(1-\alpha)+\binom{N_{0}}{i}%
\alpha]\nonumber\\
&  \overset{}{\geq}\alpha^{N_{0}}+\alpha%
{\textstyle\sum\nolimits_{i=1}^{N_{0}}}
\alpha^{N_{0}-i}(1-\alpha)^{i}\binom{N_{0}}{i}\nonumber\\
&  \overset{}{=}\alpha^{N_{0}}+\alpha(1-\alpha^{N_{0}})\nonumber\\
&  \overset{}{>}\alpha, \label{bound from other side}%
\end{align}
where the second equality uses $\delta^{0,0}=\binom{N_{1}}{0}=1,$ the first
inequality follows from setting $\delta^{i,0}=0$ for $i=1,...,N_{0},$ and the
second inequality follows from $\alpha\in(0,1).$ The remainder of the proof is
as above, namely, by continuity $P_{s_{\delta}}(Y_{B(\delta(w))}=0)>\alpha$
for a $s_{\delta}$ with $(a_{0},a_{1})=(\alpha-\varepsilon,1)$ for a small
enough $\varepsilon>0.$ For such $s_{\delta}$ regret equals 1.\medskip

Now consider the case $\alpha=0.$ Take any $\delta$ and define $s_{\delta}$ by
setting $a_{0}=P_{s_{\delta}}(Y_{0}=0)=0$ and $a_{1}=P_{s_{\delta}}%
(Y_{1}=0)=\varepsilon>0.$ Then, $q_{s_{\delta},0}(Y_{1})=0$ and any any
$q\in\lbrack0,1]$ satisfies the condition in (\ref{quantile definition}) for
$X=Y_{0}.$ Thus, with $r=1$ the $0$-quantile of $Y_{0}$ equals 1. Thus, in
order for regret to equal 1 under $s_{\delta}$ it is enough to establish that
the $0$-quantile of $Y_{B(\delta(w))}$ equals 0. The latter follows if
$P_{s_{\delta}}(Y_{B(\delta(w))}=0)>0.$ From (\ref{derivation})
\begin{equation}
P_{s_{\delta}}(Y_{B(\delta(w))}=0)=\varepsilon%
{\textstyle\sum\nolimits_{j=0}^{N_{1}}}
\varepsilon^{N_{1}-j}(1-\varepsilon)^{j}\delta^{N_{0},j} \label{alpha null.}%
\end{equation}
with $\delta^{N_{0},j}=%
{\textstyle\sum\nolimits_{v\in\{0,1\}^{N_{1}},\text{ }\overline{v}=j}}
\delta((1,...,1,v^{\prime})^{\prime}).$ If $P_{s_{\delta}}(Y_{B(\delta
(w))}=0)>0$ the proof is complete. If not, and $P_{s_{\delta}}(Y_{B(\delta
(w))}=0)=0,$ it must be that $\delta((1,...,1,v^{\prime})^{\prime})=0$ for all
$v\in\{0,1\}^{N_{1}}.$ If $\delta((1,...,1,v^{\prime})^{\prime})=0$ for all
$v\in\{0,1\}^{N_{1}}$, define $s_{\delta}^{\prime}$ by setting $a_{0}%
=P_{s_{\delta}^{\prime}}(Y_{0}=0)=\varepsilon$ and $a_{1}=P_{s_{\delta
}^{\prime}}(Y_{1}=0)=0.$ Then, by (\ref{derivation}), with $\delta^{i,N_{1}}=%
{\textstyle\sum\nolimits_{u\in\{0,1\}^{N_{0}},\overline{u}=i}}
\delta((u^{\prime},1,...,1)^{\prime})\leq\binom{N_{0}}{i}$
\begin{equation}
P_{s_{\delta}^{\prime}}(Y_{B(\delta(w))}=0)=\varepsilon%
{\textstyle\sum\nolimits_{i=0}^{N_{0}}}
\varepsilon^{N_{0}-i}(1-\varepsilon)^{i}(\binom{N_{0}}{i}-\delta^{i,N_{1}})
\label{final case to consider.}%
\end{equation}
But $1-\delta((1,...,1,v^{\prime})^{\prime})=1$ for all $v\in\{0,1\}^{N_{1}}$
implies $\binom{N_{0}}{i}-\delta^{i,N_{1}}=1$ when $i=N_{0}.$ Thus, by
(\ref{final case to consider.}), $P_{s_{\delta}^{\prime}}(Y_{B(\delta
(w))}=0)>0.$ But, the latter implies that regret equals 1 under $s_{\delta
}^{\prime}.\smallskip$

Finally, consider the case $(\alpha=1$ and $r=0).$ Consider a treatment rule
$\delta$ that is not equal to 1 wp1 and not equal to 0 wp1. We will show that
$\max_{s\in\mathbb{S}}R(\delta,s)=0.$ To see this, fix $s\in\mathbb{S}.$ By
the definition of an $\alpha$-quantile in (\ref{quantile definition}) and the
choice $r=0$ note that for any $q<q_{s,1}(Y_{0})$ it must be that $P_{s}%
(Y_{0}\leq q)<1$ and likewise for any $q<q_{s,1}(Y_{1})$ it must be that
$P_{s}(Y_{1}\leq q)<1.$ Therefore, using (\ref{prob derivation}), for any
$q<\max\{q_{s,1}(Y_{0}),q_{s,1}(Y_{1})\}$ we have%
\[
P_{s}(Y_{B(\delta(w))}\leq q)=P_{s}(B(\delta(w))=1)P_{s}(Y_{1}\leq
q)+P_{s}(B(\delta(w))=0)P_{s}(Y_{0}\leq q)<1
\]
which implies $q<q_{s,1}(Y_{B(\delta(w))}).$ By Lemma \ref{max utility}
$q_{s,1}(Y_{B(\delta(w))})\leq\max\{q_{s,1}(Y_{0}),q_{s,1}(Y_{1})\}$ and thus
$q_{s,1}(Y_{B(\delta(w))})=\max\{q_{s,1}(Y_{0}),q_{s,1}(Y_{1})\}$. But that
implies that $R(\delta,s)=0$ and because $s\in\mathbb{S}$ that proves the
desired result$.$\medskip

\textbf{Part (ii).} The proof of (ii) is similar to case (i). Again we start
off with $\alpha\in(0,1)$ and define $s_{\delta}$ as in (\ref{definition of s}%
). We obtain
\begin{align}
&  P_{s_{\delta}}(Y_{B(\delta(w))}\overset{}{=}0)\nonumber\\
&  =%
{\textstyle\sum\nolimits_{\substack{t=(t_{1},...,t_{N})^{\prime}\in
\{0,1\}^{N},\\y=(y_{t_{1}},...,y_{t_{N}})^{\prime}\in\{0,1\}^{N}}}}
p^{\overline{t}}(1-p)^{N-\overline{t}}a_{0}^{N_{0ty}}(1-a_{0})^{N-\overline
{t}-N_{0ty}}a_{1}^{N_{1ty}}(1-a_{1})^{\overline{t}-N_{1ty}}\nonumber\\
&  \times\lbrack\delta((t,y))a_{1}+(1-\delta((t,y)))a_{0}]\nonumber\\
&  =%
{\textstyle\sum\nolimits_{\substack{T=0,...,N,\\i=0,...,N-T,\\j=0,...,T}}}
p^{T}(1-p)^{N-T}a_{0}^{i}(1-a_{0})^{N-T-i}a_{1}^{j}(1-a_{1})^{T-j}\nonumber\\
&  \times\lbrack\delta^{T,i,j}a_{1}+(\binom{N}{T}\binom{N-T}{i}\binom{T}%
{j}-\delta^{T,i,j})a_{0}]\nonumber\\
&  =%
{\textstyle\sum\nolimits_{\substack{T=0,...,N,\\i=0,...,N-T,\\j=0,...,T}}}
p^{T}(1-p)^{N-T}a_{0}^{i}(1-a_{0})^{N-T-i}a_{1}^{j}(1-a_{1})^{T-j}\nonumber\\
&  \times\lbrack(a_{1}-a_{0})\delta^{T,i,j}+\binom{N}{T}\binom{N-T}{i}%
\binom{T}{j}a_{0}], \label{random assignment prob}%
\end{align}
where for given vectors $t\in\{0,1\}^{N}$ and $y\in\{0,1\}^{N},$ $N_{0ty}$
denotes the number of $y_{t_{i}},$ $i=1,...,N,$ for which $y_{t_{i}}=0$ and
$t_{i}=0,$ $N_{1ty}$ is the number of $y_{t_{i}},$ $i=1,...,N,$ for which
$y_{t_{i}}=0$ and $t_{i}=1,$ where again $\overline{t}\in\mathbb{N}$ for a
vector $t\in\{0,1\}^{N}$ denotes the sum of its components, and
\begin{equation}
\delta^{T,i,j}:=%
{\textstyle\sum\nolimits_{\substack{t=(t_{1},...,t_{N})^{\prime}\in
\{0,1\}^{N},\\y=(y_{t_{1}},...,y_{t_{N}})^{\prime}\in\{0,1\}^{N}%
,\\\overline{t}=T,\text{ }N_{0ty}=i,\text{ }N_{1ty}=j}}}
\delta((t,y)). \label{def delta}%
\end{equation}
Note that $N_{0ty}\leq N-\overline{t}$ and $N_{1ty}\leq\overline{t}.$ In the
second line of (\ref{random assignment prob}), the factor $p^{\overline{t}%
}(1-p)^{N-\overline{t}}$ captures the probability of observing exactly
$\overline{t}$ treatments and $N-\overline{t}$ controls in the sample, while
the factor $a_{0}^{N_{0ty}}(1-a_{0})^{N-\overline{t}-N_{0ty}}$ captures the
probability that among the $N-\overline{t}$ observations from the control
group exactly $N_{0ty}$ zeros are observed, while, finally, the factor
$a_{1}^{N_{1ty}}(1-a_{1})^{\overline{t}-N_{1ty}}$ captures the probability
that among the $\overline{t}$ observations from treated individuals exactly
$N_{1ty}$ zeros are observed.

Evaluating $P_{s_{\delta}}(Y_{B(\delta(w))}=0)$ when $s_{\delta}$ takes
$(a_{0},a_{1})=(1,\alpha)$ we obtain%
\begin{equation}
P_{s_{\delta}}(Y_{B(\delta(w))}=0)=%
{\textstyle\sum\nolimits_{\substack{T=0,...,N,\\j=0,...,T}}}
p^{T}(1-p)^{N-T}\alpha^{j}(1-\alpha)^{T-j}[(\alpha-1)\delta^{T,N-T,j}%
+\binom{N}{T}\binom{T}{j}]. \label{prob with part a values}%
\end{equation}
Because $(\alpha-1)<0$ and $\alpha>0$, $P_{s_{\delta}}(Y_{B(\delta(w))}=0)$ is
strictly decreasing in $\delta^{T,N-T,j}$ for all $T=0,...,N$ and $j=0,...,T$
and is minimized when $\delta^{T,N-T,j}$ takes on its maximal value $\binom
{N}{T}\binom{T}{j}$. Therefore
\begin{align}
P_{s_{\delta}}(Y_{B(\delta(w))}\overset{}{=}0)  &  \geq%
{\textstyle\sum\nolimits_{\substack{T=0,...,N,\\j=0,...,T}}}
p^{T}(1-p)^{N-T}\alpha^{j}(1-\alpha)^{T-j}\alpha\binom{N}{T}\binom{T}%
{j}\nonumber\\
&  =\alpha%
{\textstyle\sum\nolimits_{T=0,...,N}}
\binom{N}{T}p^{T}(1-p)^{N-T}%
{\textstyle\sum\nolimits_{j=0,...,T}}
\binom{T}{j}\alpha^{j}(1-\alpha)^{T-j}\nonumber\\
&  =\alpha. \label{prob in random assignment case}%
\end{align}
If $\delta^{T,N-T,j}<\binom{N}{T}\binom{T}{j}$ for any $T=0,...,N$ and
$j=0,...,T$ then by an argument as in part (ii) it follows that an $s_{\delta
}$ with $(a_{0},a_{1})=(1,\alpha-\varepsilon)$ for $\varepsilon>0$ small
enough leads to regret of 1. This holds because for that $s_{\delta}$ we have
$q_{s_{\delta},\alpha}(Y_{0})=0,$ $q_{s_{\delta},\alpha}(Y_{1})=1,$ and
$q_{s_{\delta},\alpha}(Y_{B(\delta(w))})=0.$

If on the other hand, $\delta^{T,N-T,j}=\binom{N}{T}\binom{T}{j}$ for all
$T=0,...,N$ and $j=0,...,T,$ then, as in part (ii) one can show that
$s_{\delta}$ defined by $(a_{0},a_{1})=(\alpha-\varepsilon,1)$ for a small
enough $\varepsilon>0$ leads to regret of 1. Namely, for $s_{\delta}$ with
$(a_{0},a_{1})=(\alpha,1)$ we get from (\ref{random assignment prob})
\begin{align}
&  P_{s_{\delta}}(Y_{B(\delta(w))}\overset{}{=}0)\nonumber\\
&  \overset{}{=}%
{\textstyle\sum\nolimits_{T=0,...,N}}
p^{T}(1-p)^{N-T}%
{\textstyle\sum\nolimits_{i=0,...,N-T}}
\alpha^{i}(1-\alpha)^{N-T-i}[(1-\alpha)\delta^{T,i,T}+\binom{N}{T}\binom
{N-T}{i}\alpha]\nonumber\\
&  \overset{}{=}%
{\textstyle\sum\nolimits_{T=0,...,N}}
p^{T}(1-p)^{N-T}\nonumber\\
&  \times\{\alpha^{N-T}\binom{N}{T}+%
{\textstyle\sum\nolimits_{i=0,...,N-T-1}}
\alpha^{i}(1-\alpha)^{N-T-i}[(1-\alpha)\delta^{T,i,T}+\binom{N}{T}\binom
{N-T}{i}\alpha]\}\nonumber\\
&  \overset{}{\geq}%
{\textstyle\sum\nolimits_{T=0,...,N}}
\binom{N}{T}p^{T}(1-p)^{N-T}\{\alpha^{N-T}+\alpha%
{\textstyle\sum\nolimits_{i=0,...,N-T-1}}
\binom{N-T}{i}\alpha^{i}(1-\alpha)^{N-T-i}\}\nonumber\\
&  \overset{}{=}%
{\textstyle\sum\nolimits_{T=0,...,N}}
\binom{N}{T}p^{T}(1-p)^{N-T}\{\alpha^{N-T}+\alpha(1-\alpha^{N-T})\}\nonumber\\
&  \overset{}{>}\alpha%
{\textstyle\sum\nolimits_{T=0,...,N}}
\binom{N}{T}p^{T}(1-p)^{N-T}\nonumber\\
&  \overset{}{=}\alpha, \label{final step part ii}%
\end{align}
where the second equality simply takes the last summand with index $i=N-T$ out
of the second summation sign and uses $\delta^{T,N-T,T}=\binom{N}{T}$, the
inequality is obtained by setting $\delta^{T,i,T}=0$ for all $T=0,...,N$ and
$i=0,...,N-T-1,$ and the strict inequality uses $\alpha^{N-T}+\alpha
(1-\alpha^{N-T})>\alpha.$ By an argument as in part (i),
(\ref{final step part ii}) implies that for $s_{\delta}^{\prime}$ with
$(a_{0},a_{1})=(\alpha-\varepsilon,1)$ with sufficiently small $\varepsilon>0$
regret of 1 can be obtained because under $s_{\delta}^{\prime}$ we have
$q_{s_{\delta}^{\prime},\alpha}(Y_{0})=1,$ $q_{s_{\delta}^{\prime},\alpha
}(Y_{1})=0,$ and $q_{s_{\delta}^{\prime},\alpha}(Y_{B(\delta(w))}%
)=0.$\smallskip

Next, assume $\alpha=0.$ Define $s_{\delta}$ such that $a_{0}=P_{s_{\delta}%
}(Y_{0}=0)=0$ and $a_{1}=P_{s_{\delta}}(Y_{1}=0)=\varepsilon>0.$ That and
$r=1$ imply that $q_{s_{\delta},\alpha}(Y_{0})=1$ and $q_{s_{\delta},\alpha
}(Y_{1})=0.$ Then, from (\ref{random assignment prob})%
\begin{equation}
P_{s_{\delta}}(Y_{B(\delta(w))}=0)=%
{\textstyle\sum\nolimits_{\substack{t=(t_{1},...,t_{N})^{\prime}\in
\{0,1\}^{N},\\y=(y_{t_{1}},...,y_{t_{N}})^{\prime}\in\{0,1\}^{N},\\y_{t_{j}%
}=1\text{ if }t_{j}=0\text{ for }j=1,...,N}}}
p^{\overline{t}}(1-p)^{N-\overline{t}}\varepsilon^{N_{1ty}}(1-\varepsilon
)^{\overline{t}-N_{1ty}}\delta((t,y))\varepsilon. \label{rd alpha 0}%
\end{equation}
Thus, $P_{s_{\delta}}(Y_{B(\delta(w))}=0)>0$ (which implies $R(\delta
,s_{\delta})=1$), unless $\delta((t,y))=0$ for all vectors $t,y\in\{0,1\}^{N}$
such that $y_{t_{j}}=1$ if $t_{j}=0$ for $j=1,...,N.$ For those $\delta$
consider instead, $s_{\delta}^{\prime}$ such that $a_{0}=P_{s_{\delta}%
^{\prime}}(Y_{0}=0)=\varepsilon>0$ and $a_{1}=P_{s_{\delta}^{\prime}}%
(Y_{1}=0)=0.$ Then%
\begin{equation}
P_{s_{\delta}^{\prime}}(Y_{B(\delta(w))}=0)=%
{\textstyle\sum\nolimits_{\substack{t=(t_{1},...,t_{N})^{\prime}\in
\{0,1\}^{N},\\y=(y_{t_{1}},...,y_{t_{N}})^{\prime}\in\{0,1\}^{N}\\y_{t_{j}%
}=1\text{ if }t_{j}=1\text{ for }j=1,...,N}}}
p^{\overline{t}}(1-p)^{N-\overline{t}}\varepsilon^{N_{0ty}}(1-\varepsilon
)^{N-\overline{t}-N_{0ty}}(1-\delta((t,y)))\varepsilon\label{rd alpha 0 other}%
\end{equation}
exceeds zero, because $1-\delta((0^{N\prime},1^{N^{\prime}})^{\prime})=1$ and
the vectors $t=0^{N}$ and $y=1^{N}$ appear in the sum in
(\ref{rd alpha 0 other})$.$ Thus $R(\delta,s_{\delta}^{\prime})=1.$

Finally, the case $(\alpha=1$ and $r=0)$ is handled exactly as in part
(i).$\bigskip$

\textbf{Part (iii).} We first show the following lemma$.$

\begin{lemma}
\label{lemma in prop}Let $(\alpha\in(0,1)$ and $r\in\lbrack0,1])$ or
$(\alpha=0$ and $r=1).$ If $\delta\neq0$ then there exists a $s_{\delta}%
\in\mathbb{S}$ such that $R(\delta,s_{\delta})=q_{\alpha}(Y_{0}).$
\end{lemma}

\noindent\textbf{Proof of Lemma \ref{lemma in prop}}. The case $q_{\alpha
}(Y_{0})=0$ is trivial. Simply take $s_{\delta}$ such that $P_{s_{\delta}%
}(Y_{0}=0)=P_{s_{\delta}}(Y_{1}=0)=1.$ Thus, from now on we assume $q_{\alpha
}(Y_{0})>0.$

Assume first $\alpha\in(0,1).$ Because $\delta\neq0$ there exists a vector
\begin{equation}
\widetilde{w}=(y_{1,1},...,y_{1,N})\text{ such that }\delta(\widetilde{w})>0,
\label{wtilda}%
\end{equation}
where $y_{1,i}\in\lbrack0,1]$ for $i=1,...,N.$ (When $N=0$ then we don't need
to define $\widetilde{w})$.

\emph{Definition of distributions of }$Y_{0}$\emph{ and }$Y_{1}:$ We now
define a $s_{\delta}\in\mathbb{S}$ for which $R(\delta,s_{\delta})=q_{\alpha
}(Y_{0})$ and under $s_{\delta}\in\mathbb{S}$, $Y_{0}$ and $Y_{1}$ have
independent discrete marginals$.$ Namely, $Y_{0}$ is defined by $P_{s_{\delta
}}(Y_{0}=0)=\alpha-\varepsilon$ for some small $\varepsilon>0$ $($and
obviously $\varepsilon<\alpha)$ to be further restricted below and
$P_{s_{\delta}}(Y_{0}=q_{\alpha}(Y_{0}))=1-(\alpha-\varepsilon).$ Thus
$q_{s_{\delta},\alpha}(Y_{0})=q_{\alpha}(Y_{0})$.

When $N=0$ set $P_{s_{\delta}}(Y_{1}=0)=1$ and note that $P_{s_{\delta}%
}(Y_{B(\delta)}=0)=\delta+(1-\delta)(\alpha-\varepsilon)$ which exceeds
$\alpha$ for $\varepsilon$ small enough (using $\alpha<1).$ Thus
$q_{s_{\delta},\alpha}(Y_{1})=0$ and $q_{s_{\delta},\alpha}(Y_{B(\delta)})=0$
which implies the desired result $R(\delta,s_{\delta})=q_{\alpha}(Y_{0}).$

When $N\geq1$ the discrete distribution of $Y_{1}$ under $s_{\delta}$ is
defined by
\begin{align}
P_{s_{\delta}}(Y_{1}  &  =0)=\frac{\alpha+1}{2}+\frac{1}{N}[1-\frac{\alpha
+1}{2}]%
{\textstyle\sum\nolimits_{j=1}^{N}}
I(y_{1,j}=0),\nonumber\\
P_{s_{\delta}}(Y_{1}  &  =y_{1,i})=\frac{1}{N}[1-\frac{\alpha+1}{2}]%
{\textstyle\sum\nolimits_{j=1}^{N}}
I(y_{1,j}=y_{1,i})\text{ for all }i=1,...,N\text{ s.t. }y_{1,i}\neq0,
\label{def of distribution of Y1}%
\end{align}
where the $y_{1,i}$ $i=1,...,N$ appear as the components of $\widetilde{w}$ in
(\ref{wtilda}). Note that all probabilities in
(\ref{def of distribution of Y1}) are properly defined, that is, they are all
in the interval $(0,1]$, and sum up to 1. Just for clarity $%
{\textstyle\sum\nolimits_{j=1}^{N}}
I(y_{1,j}=0)$ equals the number of zero components in the vector
$\widetilde{w}.$ Note that several of the $y_{1,i}$ for $i=1,...,N$ may be
identical and therefore $Y_{1}$ maybe supported on strictly fewer than $N+1$
points.\smallskip

With these definitions we obtain $q_{s_{\delta},\alpha}(Y_{1})=0$ (because
$.5(\alpha+1)>\alpha)$ and%
\begin{align}
&  P_{s_{\delta}}(Y_{B(\delta(w))}\overset{}{=}0)\nonumber\\
&  =%
{\textstyle\sum\nolimits_{\substack{\widehat{w}=(w_{1},...,w_{N})\\w_{i}%
\in\{0,y_{1,1},...,y_{1,N}\},\text{ }i=1,...,N}}}
{\textstyle\prod\nolimits_{i=1}^{N}}
P_{s_{\delta}}(Y_{1}=w_{i})[\delta(\widehat{w})P_{s_{\delta}}(Y_{1}%
=0)+(1-\delta(\widehat{w}))P_{s_{\delta}}(Y_{0}=0)]\nonumber\\
&  =\{%
{\textstyle\sum\nolimits_{\substack{\widehat{w}=(w_{1},...,w_{N})\\w_{i}%
\in\{0,y_{1,1},...,y_{1,N}\},\text{ }i=1,...,N}}}
{\textstyle\prod\nolimits_{i=1}^{N}}
P_{s_{\delta}}(Y_{1}=w_{i})\delta(\widehat{w})[P_{s_{\delta}}(Y_{1}%
=0)-P_{s_{\delta}}(Y_{0}=0)]\}+P_{s_{\delta}}(Y_{0}=0)\nonumber\\
&  \geq\{%
{\textstyle\prod\nolimits_{i=1}^{N}}
P_{s_{\delta}}(Y_{1}=y_{1,i})\delta(\widetilde{w})[\frac{\alpha+1}{2}%
-(\alpha-\varepsilon)]\}+\alpha-\varepsilon, \label{derivation for quantile}%
\end{align}
where the inequality is obtained by setting $\delta(\widehat{w})=0$ except for
when $\widehat{w}=\widetilde{w},$ using that $P_{s_{\delta}}(Y_{1}%
=0)-P_{s_{\delta}}(Y_{0}=0)\geq\frac{\alpha+1}{2}-(\alpha-\varepsilon)>0.$ We
claim that $P_{s_{\delta}}(Y_{B(\delta(w))}=0)>\alpha$ (and thus $q_{s,\alpha
}(Y_{B(\delta(w))})=0$) for sufficiently small $\varepsilon>0.$ But the claim
is equivalent to%
\begin{equation}%
{\textstyle\prod\nolimits_{i=1}^{N}}
P_{s_{\delta}}(Y_{1}=y_{1,i})\delta(\widetilde{w})[\frac{1-\alpha}%
{2}+\varepsilon]>\varepsilon\label{final step of iii}%
\end{equation}
which clearly holds true for $\varepsilon>0$ sufficiently small (using
$\alpha<1)$.

When $\alpha=0$ (in which case we take $r=1)$ we can use the same proof
structure except we use $s_{\delta}$ such that $P_{s_{\delta}}(Y_{0}%
=q_{\alpha}(Y_{0}))=1$ (which implies $q_{s_{\delta},\alpha}(Y_{0})=q_{\alpha
}(Y_{0}))$ and the distribution for $Y_{1}$ defined in
(\ref{def of distribution of Y1}) with $\alpha=0.$ $\square\medskip$

Note throughout that for $s\in\mathbb{S}$ we have $q_{\alpha}(Y_{0}%
)=q_{s,\alpha}(Y_{0}).$

Consider first the case $q_{\alpha}(Y_{0})>1/2.$ For $\delta^{0}\equiv0,$ we
will show next that
\begin{equation}
\max_{s\in\mathbb{S}}R(\delta^{0},s)=1-q_{\alpha}(Y_{0}).
\label{further result}%
\end{equation}
The statement in (\ref{further result}) follows easily because for any
$s\in\mathbb{S}$ for which $q_{s,\alpha}(Y_{1})=1$ we obtain $R(\delta
^{0},s)=1-q_{\alpha}(Y_{0}).$ Furthermore, because always%
\begin{equation}
q_{s,\alpha}(Y_{B(\delta(w))})\in\lbrack\min\{q_{s,\alpha}(Y_{0}),q_{s,\alpha
}(Y_{1})\},\max\{q_{s,\alpha}(Y_{0}),q_{s,\alpha}(Y_{1})\}]
\label{quantily bounds}%
\end{equation}
(as proven in Lemma \ref{max utility}(i) below) $1-q_{s,\alpha}(Y_{0})$ is the
maximal possible regret for any $s\in\mathbb{S}$ for which $q_{s,\alpha}%
(Y_{1})>q_{s,\alpha}(Y_{0}).$ For any $s\in\mathbb{S}$ such that $q_{s,\alpha
}(Y_{1})\leq q_{s,\alpha}(Y_{0})$ regret is zero. That proves
(\ref{further result}).

Because when $q_{\alpha}(Y_{0})>1/2,$ we have $1-q_{s,\alpha}(Y_{0}%
)<q_{s,\alpha}(Y_{0}),$ (\ref{further result}) combined with the lemma imply
the desired result when $q_{\alpha}(Y_{0})>1/2$.\smallskip

Next consider the case $q_{\alpha}(Y_{0})=1/2.$ Statement
(\ref{further result}) (that also holds when $q_{\alpha}(Y_{0})=1/2)$, Lemma
\ref{lemma in prop}, together with the fact that regret is bounded by 1/2 (by
(\ref{quantily bounds}) when $q_{s,\alpha}(Y_{0})=1/2),$ establish that all
treatment rules are minimax regret.

Finally, consider the case $q_{\alpha}(Y_{0})<1/2.$ We have $\max
_{s\in\mathbb{S}}R(\delta^{\ast},s)=q_{s,\alpha}(Y_{0})$ for $\delta^{\ast}%
=1$. Lemma \ref{lemma in prop} establishes that for $\delta\neq0$ $\max
_{s\in\mathbb{S}}R(\delta,s)\geq q_{s,\alpha}(Y_{0})$ while for $\delta
^{0}\equiv0,$ $\max_{s\in\mathbb{S}}R(\delta^{0},s)=1-q_{s,\alpha}%
(Y_{0})>q_{s,\alpha}(Y_{0}).$ It follows that $\delta^{\ast}=1$ is a minimax rule.

Finally, the case where $\alpha=1$ and $r=0$ is proven as in part (i).
$\square\medskip$

\textbf{Proof of Corollary
\ref{Analogue to Proposition 1 in Stoye 2009 with restrictions}(a)} This
statement follows as a corollary from the proof of Proposition
\ref{Analogue to Proposition 1 in Stoye 2009} because that proof only used
Bernoulli distributions for nature for parts (i)-(ii) and discrete
distributions in part (iii).$\medskip$

\textbf{(b) (i) }Consider an arbitrary treatment rule $\delta.$ Consider a
state of nature $s=s_{\alpha,\varepsilon,n}$ for $n\in\mathbb{N}$ such that
$Y_{0}$ has density $n$ on $[0,1/n]$ and $Y_{1}$ has density $f=f_{\alpha
,\varepsilon,n}$ which equals $n(\alpha-\varepsilon)$ on $[0,1/n]$ and
$n(1-\alpha+\varepsilon)$ on $[1-1/n,1]$ (and zero elsewhere) for some small
$\varepsilon\in\lbrack0,\alpha/2]$ to be specified more precisely below. Note
that these continuous distributions are chosen in order to approximate as
$n\rightarrow\infty$ the Bernoulli distributions used in the proof of
Proposition \ref{Analogue to Proposition 1 in Stoye 2009}(i). With these
definitions
\begin{equation}
q_{s,\alpha}(Y_{0})=\alpha/n\text{ and }q_{s,\alpha}(Y_{1})=1-1/n+\varepsilon
/(n(1-\alpha+\varepsilon)). \label{quantiles y0 and 1}%
\end{equation}

Consider an arbitrary $q\in(0,1).$\textbf{ }Then for all $n$ such that
$1/n<q<1-1/n$%
\begin{align}
&  P_{s}(Y_{B(\delta(w))}\overset{}{\leq}q)\nonumber\\
&  =%
{\textstyle\int\nolimits_{0}^{1/n}}
n...%
{\textstyle\int\nolimits_{0}^{1/n}}
n%
{\textstyle\int\nolimits_{0}^{1}}
f(y_{11})...%
{\textstyle\int\nolimits_{0}^{1}}
f(y_{1N_{1}})[\delta(y)(\alpha-\varepsilon)+(1-\delta(y))]dy_{1N_{1}%
}...dy_{11}dy_{0N_{0}}...dy_{01}\nonumber\\
&  \geq(\alpha-\varepsilon)%
{\textstyle\int\nolimits_{0}^{1/n}}
n...%
{\textstyle\int\nolimits_{0}^{1/n}}
n%
{\textstyle\int\nolimits_{0}^{1}}
f(y_{11})...%
{\textstyle\int\nolimits_{0}^{1}}
f(y_{1N_{1}})dy_{1N_{1}}...dy_{11}dy_{0N_{0}}...dy_{01}\nonumber\\
&  =(\alpha-\varepsilon), \label{continuous case}%
\end{align}
where $y:=(y_{01},...,y_{0N_{0}},y_{11},...,y_{1N_{1}})^{\prime}$ and the
inequality comes from picking $\delta^{1}\equiv1$ given that $\alpha
-\varepsilon-1<0.$

Note that if $P_{s}(\delta(w)=1)<1$ for one choice of $\varepsilon\in
\lbrack0,\alpha/2]$ then it holds for any $\varepsilon\in\lbrack0,\alpha/2].$

Case 1: $P_{s}(\delta(w)=1)<1$. By the same steps as in (\ref{continuous case}%
) with $\varepsilon=0$ we obtain $P_{s}(Y_{B(\delta(w))}\leq q)>\alpha$ and,
by continuity in $\varepsilon$, then also for $\varepsilon>0$ sufficiently
small. It follows that the $\alpha$-quantile of $Y_{B(\delta(w))}$ is
nonbigger than $q$.

Case 2: $P_{s}(\delta(w)=1)=1$. In that case, consider a state of nature
$\widetilde{s}=\widetilde{s}_{\alpha,\varepsilon,n}$ for $n\in\mathbb{N}$ such
that $Y_{1}$ has density $n$ on $[0,1/n]$ and $Y_{0}$ has density
$\widetilde{f}=\widetilde{f}_{\alpha,\varepsilon,n}$ which equals
$n(\alpha-\varepsilon)$ on $[0,1/n]$ and $n(1-\alpha+\varepsilon)$ on
$[1-1/n,1]$ (and zero elsewhere). Thus,
\begin{equation}
q_{\widetilde{s},\alpha}(Y_{1})=\alpha/n\text{ and }q_{\widetilde{s},\alpha
}(Y_{0})=1-1/n+\varepsilon/(n(1-\alpha+\varepsilon))
\label{quantiles yo and 1 again}%
\end{equation}
and, analogously to (\ref{continuous case}),%
\begin{align}
&  P_{\widetilde{s}}(Y_{B(\delta(w))}\leq q)\nonumber\\
&  =%
{\textstyle\int\nolimits_{0}^{1}}
\widetilde{f}(y_{01})...%
{\textstyle\int\nolimits_{0}^{1}}
\widetilde{f}(y_{0N_{0}})%
{\textstyle\int\nolimits_{0}^{1/n}}
n...%
{\textstyle\int\nolimits_{0}^{1/n}}
n[\delta(y)+(1-\delta(y))(\alpha-\varepsilon)]dy_{1N_{1}}...dy_{11}dy_{0N_{0}%
}...dy_{01}\nonumber\\
&  =%
{\textstyle\int\nolimits_{0}^{1/n}}
\widetilde{f}(y_{01})...%
{\textstyle\int\nolimits_{0}^{1/n}}
\widetilde{f}(y_{0N_{0}})%
{\textstyle\int\nolimits_{0}^{1/n}}
n...%
{\textstyle\int\nolimits_{0}^{1/n}}
ndy_{1N_{1}}...dy_{11}dy_{0N_{0}}...dy_{01}\nonumber\\
&  +%
{\textstyle\int}
...%
{\textstyle\int\nolimits_{A_{n}}}
\widetilde{f}(y_{01})...\widetilde{f}(y_{0N_{0}})%
{\textstyle\int\nolimits_{0}^{1/n}}
n...%
{\textstyle\int\nolimits_{0}^{1/n}}
n[\delta(y)+(1-\delta(y))(\alpha-\varepsilon)]dy_{1N_{1}}...dy_{11}dy_{0N_{0}%
}...dy_{01}\nonumber\\
&  \geq(\alpha-\varepsilon)^{N_{0}}+(\alpha-\varepsilon)%
{\textstyle\int}
...%
{\textstyle\int\nolimits_{A_{n}}}
\widetilde{f}(y_{01})...\widetilde{f}(y_{0N_{0}})dy_{0N_{0}}...dy_{01}%
\nonumber\\
&  =(\alpha-\varepsilon)^{N_{0}}+(\alpha-\varepsilon)(1-(\alpha-\varepsilon
)^{N_{0}}), \label{continuous case final step}%
\end{align}
where $A_{n}:=[0,1]^{N_{0}}\backslash\lbrack0,1/n]^{N_{0}},$ the second
equality uses that $\delta(w)=1$ a.s. on $[0,1/n]^{N}$ (under $s$ and
therefore also under $\widetilde{s}),$ and the inequality is obtained by
taking $\delta(y)=0$ on the domain of integration$.$ If $\varepsilon$ was
equal to zero, the final expression in (\ref{continuous case final step})
would be strictly larger than $\alpha.$ By continuity in $\varepsilon$, that
is then also true for sufficiently small $\varepsilon>0.$ It follows that the
$\alpha$-quantile of $Y_{B(\delta(w))}$ is nonbigger than $q$.

As $q$ can be chosen arbitrarily small in that construction it follows that
both in case 1 and case 2 regret arbitrarily close to 1 can be inflicted by
nature given (\ref{quantiles y0 and 1}) and (\ref{quantiles yo and 1 again}).

Parts (ii) and (iii) follow from similar constructions. $\square$ $\medskip$

\textbf{Proof of the statements in Comment 3 below Corollary
\ref{Analogue to Proposition 1 in Stoye 2009 with restrictions}. }Note first
that $q(\delta)$ indeed exists because cdfs are continuous from the left. E.g.
$q(0)=q_{\alpha}(Y_{0})$ and $q(1)=0.$ Also note that for any $q\in
\lbrack0,1]$ the function $g(\delta):=\delta+(1-\delta)F_{Y_{0}}(q)$ is weakly
increasing in $\delta.$ It follows that $q(\delta)$ is weakly decreasing in
$\delta.$ Therefore, $q_{\alpha}(Y_{0})-q(\delta)$ is weakly increasing in
$\delta.$

If $\delta<1$ then regret of $1-q_{\alpha}(Y_{0})$ can be achieved by nature
by defining the distribution of $Y_{1}$ as $P_{s}(Y_{1}=0)=\alpha-\varepsilon$
and $P_{s}(Y_{1}=1)=1-(\alpha-\varepsilon)$ for some small $\varepsilon>0$ (to
be specified further below). Namely, with that definition, for $q_{\alpha
}(Y_{0})<1$ we have
\begin{equation}
P_{s}(Y_{B(\delta)}\leq q_{\alpha}(Y_{0}))=\delta(\alpha-\varepsilon
)+(1-\delta)F_{Y_{0}}(q_{\alpha}(Y_{0})) \label{derivation step}%
\end{equation}
which exceeds $\alpha$ for $\varepsilon$ small enough because $F_{Y_{0}%
}(q_{\alpha}(Y_{0}))>\alpha$ and $\delta<1.$ And of course $P_{s}%
(Y_{B(\delta)}\leq q_{\alpha}(Y_{0}))\geq\alpha$ when $q_{\alpha}(Y_{0})=1.$
Because the $\alpha$-quantile of $Y_{B(\delta)}$ can clearly not be smaller
than $q_{\alpha}(Y_{0})$ it follows that it is equal to $q_{\alpha}(Y_{0}).$

But then, in the case $q_{\alpha}(Y_{0})<1/2,$ it follows that for any
$\delta\in\lbrack0,1),$ $\max_{s\in\mathbb{S}}R(\delta,s)\geq1-q_{\alpha
}(Y_{0})>q_{\alpha}(Y_{0}).$ Given that for $\delta^{1}\equiv1$ maximal regret
equals $q_{\alpha}(Y_{0})$ it is the unique minimax regret rule. When
$q_{\alpha}(Y_{0})=1/2$ regret cannot exceed 1/2 and given regret of
$1-q_{\alpha}(Y_{0})$ can always be achieved by the argument above (for
$\delta<1$ and for $q_{\alpha}(Y_{0})=1/2$ also for $\delta=1)$ it follows
that all rules are minimax regret.

Furthermore, for any $\delta\in\lbrack0,1]$ regret of $q_{\alpha}%
(Y_{0})-q(\delta)$ can be achieved by nature by defining $s,$ the distribution
of $Y_{1},$ by $P_{s}(Y_{1}=0)=1.$ Then $P_{s}(Y_{B(\delta)}\leq
q(\delta))=\delta+(1-\delta)F_{Y_{0}}(q(\delta))$ which by (\ref{condition})
and $r=0$ implies $q_{s,\alpha}(Y_{B(\delta)})=q(\delta).$ By construction, no
bigger regret than $q_{\alpha}(Y_{0})-q(\delta)$ can be achieved by nature for
setups where $q_{\alpha}(Y_{1})\leq q_{\alpha}(Y_{0}).$ That proves the claim
in the case where $q_{\alpha}(Y_{0})>1/2.$ $\square\bigskip$

\textbf{Proof of Corollary \ref{statement with covariates}. Part (i). }Let
$\delta$ be an arbitrary treatment rule. We will construct a state of nature
$s_{\delta}$ that yields regret equal to 1. Consider the case where all
marginals under $s_{\delta}$ are independent Bernoulli distributions and
define
\begin{equation}
a_{0x_{k}}=P_{s_{\delta}}(Y_{0,x_{k}}=0)\text{ and }a_{1x_{k}}=P_{s_{\delta}%
}(Y_{1,x_{k}}=0) \label{def of bernoullis}%
\end{equation}
for $k=1,...,K$ (and, of course, $1-a_{0x_{k}}=P_{s_{\delta}}(Y_{0,x_{k}}=1)$
and $1-a_{1x_{k}}=P_{s_{\delta}}(Y_{1,x_{k}}=1)$)$.$ Notationwise, we index
probabilities by those distributions that they depend on. Then,%
\begin{equation}
P_{s_{\delta},F_{X}}(Y_{B(\delta_{X}(w)),X}=0)=%
{\textstyle\sum\nolimits_{k=1}^{K}}
P_{F_{X}}(X=x_{k})P_{s_{\delta},F_{X}}(Y_{B(\delta_{x_{k}}(w)),x_{k}}=0)
\label{overall probability}%
\end{equation}
and%
\begin{align}
&  P_{s_{\delta},F_{X}}(Y_{B(\delta_{x_{k}}(w)),x_{k}}\overset{}{=}%
0)\nonumber\\
&  \overset{}{=}%
{\textstyle\sum\nolimits_{\substack{\widetilde{x}=(\widetilde{x}%
_{1},...,\widetilde{x}_{N})\in\mathcal{X}^{N}\\y=(y_{t_{1}},...,y_{t_{N}}%
)\in\{0,1\}^{N}}}}
{\textstyle\prod\nolimits_{l=1}^{N_{0}}}
P_{F_{X}}(X\overset{}{=}\widetilde{x}_{l})a_{0\widetilde{x}_{l}}^{1-y_{t_{l}}%
}(1-a_{0\widetilde{x}_{l}})^{y_{t_{l}}}%
{\textstyle\prod\nolimits_{l^{\prime}=N_{0}+1}^{N}}
P_{F_{X}}(X\overset{}{=}\widetilde{x}_{l^{\prime}})a_{1\widetilde{x}%
_{l^{\prime}}}^{1-y_{t_{l^{\prime}}}}(1-a_{1\widetilde{x}_{l^{\prime}}%
})^{y_{t_{l^{\prime}}}}\nonumber\\
&  \overset{}{\times}[\delta_{x_{k}}(\widetilde{w}(\widetilde{x},y))a_{1x_{k}%
}+(1-\delta_{x_{k}}(\widetilde{w}(\widetilde{x},y)))a_{0x_{k}}],
\label{k with covariates}%
\end{align}
where in the last line $\widetilde{w}(\widetilde{x},y)$ denotes the sample
associated with $\widetilde{x}$ and $y$. Note that expressions like
$P_{s_{\delta},F_{X}}(Y_{B(\delta_{x_{k}}(w)),x_{k}}=0)$ are still indexed by
$F_{X}$ because the distribution of $w$ depends on $F_{X}.$ In the case
$(a_{0x_{k}},a_{1x_{k}})=(1,\alpha)$ for all $k=1,...,K$
(\ref{k with covariates}) becomes%
\begin{align}
&  P_{s_{\delta},F_{X}}(Y_{B(\delta_{x_{k}}(w)),x_{k}}\overset{}{=}%
0)\nonumber\\
&  \overset{}{=}%
{\textstyle\sum\nolimits_{\substack{\widetilde{x}=(\widetilde{x}%
_{1},...,\widetilde{x}_{N})\in\mathcal{X}^{N}\\y=(0,...,0,y_{t_{N_{0}+1}%
},...,y_{t_{N}})\in\{0\}^{N_{0}}\times\{0,1\}^{N_{1}}}}}
{\textstyle\prod\nolimits_{l=1}^{N_{0}}}
P_{F_{X}}(X\overset{}{=}\widetilde{x}_{l})%
{\textstyle\prod\nolimits_{l^{\prime}=N_{0}+1}^{N}}
P_{F_{X}}(X\overset{}{=}\widetilde{x}_{l^{\prime}})\alpha^{1-y_{t_{l^{\prime}%
}}}(1-\alpha)^{y_{t_{l^{\prime}}}}\nonumber\\
&  \overset{}{\times}[(\alpha-1)\delta_{x_{k}}(\widetilde{w}(\widetilde{x}%
,y))+1]\nonumber\\
&  \overset{}{\geq}[%
{\textstyle\sum\nolimits_{\substack{\widetilde{x}=(\widetilde{x}%
_{1},...,\widetilde{x}_{N})\in\mathcal{X}^{N}\\(y_{t_{N_{0}+1}},...,y_{t_{N}%
})\in\{0,1\}^{N-N_{0}}}}}
{\textstyle\prod\nolimits_{l=1}^{N_{0}}}
P_{F_{X}}(X\overset{}{=}\widetilde{x}_{l})%
{\textstyle\prod\nolimits_{l^{\prime}=N_{0}+1}^{N}}
P_{F_{X}}(X\overset{}{=}\widetilde{x}_{l^{\prime}})\alpha^{1-y_{t_{l^{\prime}%
}}}(1-\alpha)^{y_{t_{l^{\prime}}}}]\alpha\nonumber\\
&  \overset{}{=}\alpha, \label{with covariates specific}%
\end{align}
where the inequality follows from taking $\delta_{x_{k}}(\widetilde{w}%
(\widetilde{x},y))=1.$

If for some $k\in\{1,...,K\}$ $\delta_{x_{k}}(\widetilde{w}(\widetilde{x}%
,y))<1$ for some $\widetilde{w}(\widetilde{x},y)$ (with $\widetilde{x}%
=(\widetilde{x}_{1},...,\widetilde{x}_{N})\in\mathcal{X}^{N}$ and
$y=(0,...,0,y_{t_{N_{0}+1}},...,y_{t_{N}})$ with $y_{t_{j}}\in\{0,1\}$ for
$j=N_{0}+1,...,N)$ it follows that $P_{s_{\delta},F_{X}}(Y_{B(\delta_{x_{k}%
}(w)),x_{k}}=0)>\alpha$ and by (\ref{overall probability}) therefore also
$P_{s_{\delta},F_{X}}(Y_{B(\delta_{X}(w)),X}=0)>\alpha.$ Notice that
$P_{s_{\delta},F_{X}}(Y_{B(\delta_{X}(w)),X}=0)$ is continuous in $a_{1x_{k}}$
for $k=1,...,K$ and therefore $P_{s_{\delta},F_{X}}(Y_{B(\delta_{X}%
(w)),X}=0)>\alpha$ when instead of $(a_{0x_{k}},a_{1x_{k}})=(1,\alpha)$ for
$k=1,...,K,$ $s_{\delta}$ is defined by $(a_{0x_{k}},a_{1x_{k}})=(1,\alpha
-\varepsilon)$ for $k=1,...,K$ for some small enough $\varepsilon>0.$ But then
for that choice of $s_{\delta},$ regret equals 1 (as can be seen by comparing
to the quantile obtained for the treatment rule $\delta\equiv1)$ as desired.

If instead for all $k\in\{1,...,K\}$ $\delta_{x_{k}}(\widetilde{w}%
(\widetilde{x},y))=1$ for all $\widetilde{w}(\widetilde{x},y)$ (with
$\widetilde{x}=(\widetilde{x}_{1},...,\widetilde{x}_{N})\in\mathcal{X}^{N}$
and $y=(0,...,0,y_{t_{N_{0}+1}},...,y_{t_{N}})$ with $y_{t_{j}}\in\{0,1\}$ for
$j=N_{0}+1,...,N)$ nature can cause regret of 1 by using distributions defined
by $(a_{0x_{k}},a_{1x_{k}})=(\alpha-\varepsilon,1)$ for some small enough
$\varepsilon>0.$ Namely, note first that for $s_{\delta}$ defined by
$(a_{0x_{k}},a_{1x_{k}})=(\alpha,1)$ for $k=1,...,K$ we obtain from
(\ref{k with covariates})%
\begin{align}
&  P_{s_{\delta},F_{X}}(Y_{B(\delta_{x_{k}}(w)),x_{k}}\overset{}{=}%
0)\nonumber\\
&  \overset{}{=}%
{\textstyle\sum\nolimits_{\substack{\widetilde{x}=(\widetilde{x}%
_{1},...,\widetilde{x}_{N})\in\mathcal{X}^{N}\\y=(y_{t_{1}},...,y_{t_{N_{0}}%
},0,...,0)\in\{0,1\}^{N_{0}}\times\{0\}^{N_{1}}}}}
{\textstyle\prod\nolimits_{l=1}^{N_{0}}}
P_{F_{X}}(X\overset{}{=}\widetilde{x}_{l})\alpha^{1-y_{t_{l}}}(1-\alpha
)^{y_{t_{l}}}%
{\textstyle\prod\nolimits_{l^{\prime}=N_{0}+1}^{N}}
P_{F_{X}}(X\overset{}{=}\widetilde{x}_{l^{\prime}})\nonumber\\
&  \times\lbrack\delta_{x_{k}}(\widetilde{w}(\widetilde{x},y))(1-\alpha
)+\alpha]\nonumber\\
&  \overset{}{\geq}\alpha^{N_{0}}%
{\textstyle\sum\nolimits_{\widetilde{x}=(\widetilde{x}_{1},...,\widetilde{x}%
_{N})\in\mathcal{X}^{N}}}
{\textstyle\prod\nolimits_{l=1}^{N}}
P_{F_{X}}(X\overset{}{=}\widetilde{x}_{l})\nonumber\\
&  +\alpha%
{\textstyle\sum\nolimits_{\substack{\widetilde{x}=(\widetilde{x}%
_{1},...,\widetilde{x}_{N})\in\mathcal{X}^{N}\\y=(y_{t_{1}},...,y_{t_{N_{0}}%
},0,...,0)\in\{0,1\}^{N_{0}}\times\{0\}^{N_{1}},y\neq0^{N}}}}
{\textstyle\prod\nolimits_{l=1}^{N_{0}}}
P_{F_{X}}(X\overset{}{=}\widetilde{x}_{l})\alpha^{1-y_{t_{l}}}(1-\alpha
)^{y_{t_{l}}}(%
{\textstyle\prod\nolimits_{l^{\prime}=N_{0}+1}^{N}}
P_{F_{X}}(X\overset{}{=}\widetilde{x}_{l^{\prime}}))\nonumber\\
&  \overset{}{=}\alpha^{N_{0}}+\alpha-\alpha\lbrack%
{\textstyle\sum\nolimits_{\widetilde{x}=(\widetilde{x}_{1},...,\widetilde{x}%
_{N})\in\mathcal{X}^{N}}}
{\textstyle\prod\nolimits_{l=1}^{N_{0}}}
(P_{F_{X}}(X\overset{}{=}\widetilde{x}_{l})\alpha)(%
{\textstyle\prod\nolimits_{l^{\prime}=N_{0}+1}^{N}}
P_{F_{X}}(X\overset{}{=}\widetilde{x}_{l^{\prime}}))]\nonumber\\
&  \overset{}{=}\alpha^{N_{0}}+\alpha-\alpha^{N_{0}+1}, \label{last step}%
\end{align}
where the inequality follows from setting $\delta_{x_{k}}(\widetilde{w}%
(\widetilde{x},y))=0$ for all elements in the sum except for those when
$y=0^{N}$ in which case we use $\delta_{x_{k}}(\widetilde{w}(\widetilde{x}%
,y))=1.$ The second to last equality follows by first considering a sum over
all $y=(y_{t_{1}},...,y_{t_{N_{0}}},0,...,0)\in\{0,1\}^{N_{0}}\times
\{0\}^{N_{1}}$ in the second sum and then subtracting the summand associated
with $y=0^{N}.$ The final expression is strictly larger than $\alpha$ which by
an argument as above allows us to conclude that $P_{s_{\delta},F_{X}%
}(Y_{B(\delta_{x_{k}}(w)),x_{k}}=0)>\alpha$ when $s_{\delta}$ is defined by
$(a_{0x_{k}},a_{1x_{k}})=(\alpha-\varepsilon,1)$ for $k=1,...,K$ for some
small enough $\varepsilon>0.$

The proof of \textbf{part (ii)} is very similar to part (i) and omitted.

\textbf{Part (iii).} The key input of the proof is the following lemma.

\begin{lemma}
\label{lemma in corr to prop} For any treatment rule $\delta\in\mathbb{D}$
such that $\delta\neq0$ there exists an $s_{\delta}\in\mathbb{S}$ such that
$R(\delta,s_{\delta})=q_{\alpha}(Y_{0,X}).$
\end{lemma}

\textbf{Proof of Lemma \ref{lemma in corr to prop}.} We define an $s_{\delta
}\in\mathbb{S}$ by letting all marginals $(Y_{t,x_{k}})$ be independent and
discrete, namely for an $\varepsilon>0$ (to be specified later) let
$P_{s_{\delta}}(Y_{0,x_{k}}=0)=\alpha-\varepsilon$ and $P_{s_{\delta}%
}(Y_{0,x_{k}}=q_{\alpha}(Y_{0,X}))=1-(\alpha-\varepsilon)$ for every $x_{k}%
\in\mathcal{X}$. Because $P_{s_{\delta},F_{X}}(Y_{0,X}=0)=\sum_{k=1}%
^{K}P_{F_{X}}(X=x_{k})P_{s_{\delta}}(Y_{0,x_{k}}=0)=\alpha-\varepsilon$, it
follows that $q_{s,F_{X},\alpha}(Y_{0,X})=q_{\alpha}(Y_{0,X})$ as required.

For $N=0$, complete the definition of $s_{\delta}$ by setting $P_{s_{\delta}%
}(Y_{1,x_{k}}=0)=1$ for every $x_{k}\in\mathcal{X}$. Wlog assume
$\delta_{x_{l}}>0.$ Note that $P_{s_{\delta}}(Y_{B(\delta_{x_{l}}),x_{l}%
}=0)=\delta_{x_{l}}+(1-\delta_{x_{l}})(\alpha-\varepsilon)$ which, given that
$\delta_{x_{l}}>0,$ is strictly larger than $\alpha$ when $\varepsilon>0$ is
chosen sufficiently small. For every $x_{k}\in\mathcal{X}$ such that
$x_{k}\neq x_{l}$ we have $P_{s_{\delta}}(Y_{B(\delta_{x_{k}}),x_{k}}%
=0)\geq\alpha-\varepsilon$. Therefore,%
\begin{align}
&  P_{s_{\delta},F_{X}}(Y_{B(\delta_{X}),X}\overset{}{=}0)\nonumber\\
&  \overset{}{=}\sum_{k=1}^{K}P_{F_{X}}(X\overset{}{=}x_{k})P_{s_{\delta}%
}(Y_{B(\delta_{x_{k}}),x_{k}}\overset{}{=}0)\nonumber\\
&  \overset{}{\geq}P_{F_{X}}(X\overset{}{=}x_{l})(\delta_{x_{l}}%
+(1-\delta_{x_{l}})(\alpha-\varepsilon))+\sum_{k=1,k\neq l}^{K}P_{F_{X}%
}(X\overset{}{=}x_{k})(\alpha-\varepsilon)\nonumber\\
&  \overset{}{=}P_{F_{X}}(X\overset{}{=}x_{l})\delta_{x_{l}}(1-(\alpha
-\varepsilon))+(\alpha-\varepsilon)\nonumber\\
&  \overset{}{>}\alpha, \label{N equ to zero case}%
\end{align}
where the second equality follows by including the summand $P_{F_{X}}%
(X=x_{l})(\alpha-\varepsilon)$ into the summation sign, and the inequality
holds for $\varepsilon>0$ chosen small enough.\smallskip

Next consider the case $N>0$. Because $\delta\neq0$ there exists some vector
$\widetilde{w}=((\widetilde{x}_{1},\widetilde{y}_{1}),...,(\widetilde{x}%
_{N},\widetilde{y}_{N}))$ with $(\widetilde{x}_{i},\widetilde{y}_{i}%
)\in\mathcal{X}\times\lbrack0,1]$ for every $i=1,...,N$ and some covariate
$x_{l}\in\mathcal{X}$ such that $\delta_{x_{l}}(\Tilde{w})>0$. Similarly to
the proof of Proposition \ref{Analogue to Proposition 1 in Stoye 2009}(iii),
for every $x_{k}\in\mathcal{X}$ define the distribution of $Y_{1,x_{k}}$ under
$s_{\delta}$ as
\begin{align}
P_{s_{\delta}}(Y_{1,x_{k}}\overset{}{=}0)  &  =\frac{\alpha+1}{2}+\frac{1}%
{N}[1-\frac{\alpha+1}{2}]\sum_{j=1}^{N}I(\widetilde{y}_{j}=0),\nonumber\\
P_{s_{\delta}}(Y_{1,x_{k}}\overset{}{=}\widetilde{y}_{i})  &  =\frac{1}%
{N}[1-\frac{\alpha+1}{2}]\sum_{j=1}^{N}I(\widetilde{y}_{j}=\widetilde{y}%
_{i})\text{ for $i=1,...,N$ s.t. }\widetilde{\text{$y$}}\text{$_{i}\neq0.$}
\label{def y1 again}%
\end{align}
Because $P_{s_{\delta},F_{X}}(Y_{1,X}=0)=\sum_{k=1}^{K}P_{F_{X}}%
(X=x_{k})P_{s_{\delta}}(Y_{1,x_{k}}=0)$, it then follows that $q_{s,F_{x}%
,\alpha}(Y_{1,X})=0$. Now, focus on $P_{s_{\delta,F_{X}}}(Y_{B(\delta_{x_{l}%
}(w)),x_{l}}=0)$. Furthermore,%
\begin{align}
&  P_{s_{\delta,F_{X}}}(Y_{B(\delta_{x_{l}}(w)),x_{l}}\overset{}{=}%
0)\nonumber\\
&  =\sum_{\substack{w=((a_{1},b_{1}),...,(a_{N},b_{N}));\\a_{i}\in
\mathcal{X};\hspace{0.1cm}b_{i}\in\{0,\widetilde{y}_{1},...,\widetilde{y}%
_{N}\},\text{ }\\i=1,...,N}}\prod_{i=1}^{N}P_{s_{\delta},F_{x}}(Y_{1,a_{i}%
}=b_{i},X=a_{i})(\delta_{x_{l}}(w)P_{s_{\delta}}(Y_{1,x_{l}}=0)+(1-\delta
_{x_{l}}(w))P_{s_{\delta}}(Y_{0,x_{l}}=0))\nonumber\\
&  =P_{s_{\delta}}(Y_{0,x_{l}}=0)+\sum_{\substack{w=((a_{1},b_{1}%
),...,(a_{N},b_{N}));\\a_{i}\in\mathcal{X};\hspace{0.1cm}b_{i}\in
\{0,\widetilde{y}_{1},...,\widetilde{y}_{N}\},\text{ }\\i=1,...,N}}\prod
_{i=1}^{N}P_{s_{\delta},F_{x}}(Y_{1,a_{i}}=b_{i},X=a_{i})(\delta_{x_{l}%
}(w)(P_{s_{\delta}}(Y_{1,x_{l}}=0)-P_{s_{\delta}}(Y_{0,x_{l}}=0))\nonumber\\
&  \geq\alpha-\varepsilon+\prod_{i=1}^{N}P_{s_{\delta}}(Y_{1,\widetilde{x}%
_{i}}=\widetilde{y}_{i})P_{F_{X}}(X=\widetilde{x}_{i})\delta_{x_{l}}(\Tilde
{w})(\frac{\alpha+1}{2}-(\alpha-\varepsilon))\nonumber\\
&  >\alpha, \label{outcome prob}%
\end{align}
where the first inequality follows by replacing $\delta_{x_{l}}(w)$ by zero
for every $w\neq\Tilde{w}$ and the last inequality holds for small enough
$\varepsilon>0.$ By the same derivations, for any other $x_{k}\neq x_{l}$ it
follows that $P_{s_{\delta},F_{X}}(Y_{B(\delta_{x_{k}}(w)),x_{k}}=0)\geq
\alpha-\varepsilon$. Therefore, for sufficiently small $\varepsilon>0$ we have%
\begin{align}
&  P_{s_{\delta},F_{x}}(Y_{B(\delta_{X}(w)),X}\overset{}{=}0)\nonumber\\
&  =\sum_{k=1}^{K}P_{F_{X}}(X=x_{k})P_{s_{\delta}}(Y_{B(\delta_{x_{k}%
}(w)),x_{k}}=0)\nonumber\\
&  \geq\alpha-\varepsilon+P_{F_{X}}(X=x_{l})\prod_{i=1}^{N}P_{s_{\delta}%
}(Y_{1,\widetilde{x}_{i}}=\widetilde{y}_{i})P_{F_{X}}(X=\widetilde{x}%
_{i})\delta_{x_{l}}(\Tilde{w})(\frac{\alpha+1}{2}-(\alpha-\varepsilon
))\nonumber\\
&  >\alpha, \label{final step}%
\end{align}
where for the inequality we use $P_{s_{\delta}}(Y_{B(\delta_{x_{k}}(w)),x_{k}%
}=0)\geq\alpha-\varepsilon$ and the second to last line in (\ref{outcome prob}%
). Therefore, we have $q_{s,F_{X},\alpha}(Y_{0,X})=q_{\alpha}(Y_{0,X})$ and
$q_{s,F_{X},\alpha}(Y_{1,X})=q_{s,F_{X},\alpha}(Y_{B(\delta_{X}(w)),X})=0$.
Hence, the statement of the lemma follows. $\square$

Given Lemma \ref{lemma in corr to prop} the remainder of the proof of part
(iii) is exactly the same as the last part of the proof of Proposition
\ref{Analogue to Proposition 1 in Stoye 2009}(iii). The only detail to think
about is establishing that
\begin{equation}
q_{s,F_{X},\alpha}(Y_{B(\delta_{X}(w)),X})\in\lbrack min\{q_{s,F_{X},\alpha
}(Y_{0,X}),q_{s,F_{X},\alpha}(Y_{1,X})\},max\{q_{s,F_{X},\alpha}%
(Y_{0,X}),q_{s,F_{X},\alpha}(Y_{1,X})\}]. \label{quantile bounds}%
\end{equation}
The proof of that statement is analogous to the proof of Lemma
\ref{max utility} and therefore omitted. $\square$\bigskip

The following is a technical lemma that is used in the proof of Proposition
\ref{Analogue to Proposition 1 in Stoye 2009} above. It states that
$1(q_{s,\alpha}(Y_{1})\geq q_{s,\alpha}(Y_{0}))$ is an infeasible optimal rule.

\begin{lemma}
\label{max utility} For any given $r\in\lbrack0,1]$ and arbitrary
$s\in\mathbb{S}$ we have $($i$)$
\[
q_{s,\alpha}(Y_{B(\delta(w))})\in\lbrack\min\{q_{s,\alpha}(Y_{0}),q_{s,\alpha
}(Y_{1})\},\max\{q_{s,\alpha}(Y_{0}),q_{s,\alpha}(Y_{1})\}]
\]
and $($ii$)$
\[
\max_{\delta\in\mathbb{D}}u(\delta,s)=\max_{\delta\in\mathbb{D}}q_{s,\alpha
}(Y_{B(\delta(w))})=\max\{q_{s,\alpha}(Y_{0}),q_{s,\alpha}(Y_{1})\}.
\]

\end{lemma}

\noindent\textbf{Proof of Lemma \ref{max utility}. }(i) Denote by
$\overline{q}_{0,\alpha},$ $\overline{q}_{1,\alpha},$ and $\overline
{q}_{\delta,\alpha}$ the largest $\alpha$-quantile (meaning employing $r=1$ in
definition (\ref{definition of the alpha quantile})) of $Y_{0},Y_{1},$ and
$Y_{B(\delta(w))}$, respectively. Likewise, denote by \underline{$q$%
}$_{0,\alpha},$ \underline{$q$}$_{1,\alpha},$ and \underline{$q$}%
$_{\delta,\alpha}$ the smallest $\alpha$-quantile (meaning employing $r=0$ in
definition (\ref{definition of the alpha quantile})) of $Y_{0},Y_{1},$ and
$Y_{B(\delta(w))}$, respectively. As in (\ref{prob derivation})
\begin{equation}
P_{s}(Y_{B(\delta(w))}\geq q)=P_{s}(B(\delta(w))=1)P_{s}(Y_{1}\geq
q)+P_{s}(B(\delta(w))=0)P_{s}(Y_{0}\geq q). \label{repeat result}%
\end{equation}
We show first that
\begin{equation}
\overline{q}_{\delta,\alpha}\in\lbrack\min\{\overline{q}_{0,\alpha}%
,\overline{q}_{1,\alpha}\},\max\{\overline{q}_{0,\alpha},\overline
{q}_{1,\alpha}\}]. \label{claim 1}%
\end{equation}

Indeed, if $\overline{q}_{\delta,\alpha}>\max\{\overline{q}_{0,\alpha
},\overline{q}_{1,\alpha}\}$ then, from (\ref{repeat result})
\begin{align}
&  P_{s}(Y_{B(\delta(w))}\overset{}{\geq}\overline{q}_{\delta,\alpha
})\nonumber\\
&  \overset{}{=}P_{s}(B(\delta(w))\overset{}{=}1)P_{s}(Y_{1}\overset{}{\geq
}\overline{q}_{\delta,\alpha})+P_{s}(B(\delta(w))\overset{}{=}0)P_{s}%
(Y_{0}\overset{}{\geq}\overline{q}_{\delta,\alpha})\nonumber\\
&  \overset{}{<}1-\alpha,
\end{align}
a contradiction to the definition of an $\alpha$-quantile. On the other hand,
if $\overline{q}_{\delta,\alpha}<\min\{\overline{q}_{0,\alpha},\overline
{q}_{1,\alpha}\}$ then there exists $\varepsilon>0$ such that $\overline
{q}_{\delta,\alpha}+\varepsilon<\min\{\overline{q}_{0,\alpha},\overline
{q}_{1,\alpha}\}.$ But then
\begin{align}
&  P_{s}(Y_{B(\delta(w))}\overset{}{\geq}\overline{q}_{\delta,\alpha
}+\varepsilon)\nonumber\\
&  \overset{}{=}P_{s}(B(\delta(w))\overset{}{=}1)P_{s}(Y_{1}\overset{}{\geq
}\overline{q}_{\delta,\alpha}+\varepsilon)+P_{s}(B(\delta(w))\overset{}{=}%
0)P_{s}(Y_{0}\overset{}{\geq}\overline{q}_{\delta,\alpha}+\varepsilon
)\nonumber\\
&  \overset{}{\leq}P_{s}(B(\delta(w))\overset{}{=}1)P_{s}(Y_{1}\overset{}{\geq
}\overline{q}_{1,\alpha})+P_{s}(B(\delta(w))\overset{}{=}0)P_{s}%
(Y_{0}\overset{}{\geq}\overline{q}_{0,\alpha})\nonumber\\
&  \overset{}{=}1-\alpha,
\end{align}
contradicting the fact that $\overline{q}_{\delta,\alpha}$ is the largest
$\alpha$-quantile of $Y_{B(\delta(w))}.$

By an analogous argument it follows that
\begin{equation}
\underline{q}_{\delta,\alpha}\in\lbrack\min\{\underline{q}_{0,\alpha
},\underline{q}_{1,\alpha}\},\max\{\underline{q}_{0,\alpha},\underline{q}%
_{1,\alpha}\}]. \label{claim 2}%
\end{equation}
\bigskip

Define the interval $I=[\min\{\underline{q}_{0,\alpha},\underline{q}%
_{1,\alpha}\},\max\{\overline{q}_{0,\alpha},\overline{q}_{1,\alpha}\}].$

If $I=[\underline{q}_{i,\alpha},\overline{q}_{j,\alpha}]$ for $i,j\in\{0,1\},$
$i\neq j$ then
\begin{equation}
q_{s,\alpha}(Y_{B(\delta(w))})=r\overline{q}_{\delta,\alpha}%
+(1-r)\underline{q}_{\delta,\alpha} \label{conv comb}%
\end{equation}
and thus by (\ref{claim 1}) and (\ref{claim 2})
\begin{align}
q_{s,\alpha}(Y_{B(\delta(w))})  &  \leq r\overline{q}_{j,\alpha}%
+(1-r)\underline{q}_{j,\alpha}=q_{s,\alpha}(Y_{j})\leq\max\{q_{s,\alpha}%
(Y_{0}),q_{s,\alpha}(Y_{1})\}\text{ and}\nonumber\\
q_{s,\alpha}(Y_{B(\delta(w))})  &  \geq r\overline{q}_{i,\alpha}%
+(1-r)\underline{q}_{i,\alpha}=q_{s,\alpha}(Y_{i})\leq\max\{q_{s,\alpha}%
(Y_{0}),q_{s,\alpha}(Y_{1})\}. \label{bounds}%
\end{align}
For the remaining case we have $I=[\underline{q}_{i,\alpha},\overline
{q}_{i,\alpha}]$ for a $i\in\{0,1\}.$ By (\ref{claim 1}) and (\ref{claim 2})
it follows that $[\underline{q}_{\delta,\alpha},\overline{q}_{\delta,\alpha
}]\subset I$ and $[\underline{q}_{j,\alpha},\overline{q}_{j,\alpha}%
]\subset\lbrack\underline{q}_{\delta,\alpha},\overline{q}_{\delta,\alpha}].$

If $P_{s}(B(\delta(w))=i)=1$ then $Y_{B(\delta(w))}=Y_{i}$ wp1 and the lemma
trivially holds. If $P_{s}(B(\delta(w))=i)<1$, we must have $[\underline{q}%
_{j,\alpha},\overline{q}_{j,\alpha}]=[\underline{q}_{\delta,\alpha}%
,\overline{q}_{\delta,\alpha}].$ If not, if e.g., $\overline{q}_{j,\alpha
}<\overline{q}_{\delta,\alpha}$ then $P_{s}(Y_{j}\geq\overline{q}%
_{\delta,\alpha})<1-\alpha$ and $P_{s}(Y_{i}\geq\overline{q}_{\delta,\alpha
})=1-\alpha.$ But then, by (\ref{repeat result}) we get the contradiction
\begin{equation}
P_{s}(Y_{B(\delta(w))}\geq\overline{q}_{\delta,\alpha})=P_{s}(B(\delta
(w))=1)P_{s}(Y_{1}\geq\overline{q}_{\delta,\alpha})+P_{s}(B(\delta
(w))=0)P_{s}(Y_{0}\geq\overline{q}_{\delta,\alpha})<1-\alpha, \label{inequ}%
\end{equation}
where the last inequality uses $P_{s}(B(\delta(w))=i)<1.$ But if
$[\underline{q}_{j,\alpha},\overline{q}_{j,\alpha}]=[\underline{q}%
_{\delta,\alpha},\overline{q}_{\delta,\alpha}]$ then $q_{s,\alpha}%
(Y_{B(\delta(w))})=q_{s,\alpha}(Y_{j})$ and the claim in the lemma
holds.$\smallskip$

(ii)\textbf{ }Considering the two rules $\delta^{1}\equiv1$ or $\delta
^{0}\equiv0$ (that is, the rules that pick 1 (or 0) with probability 1), it
follows that $\max_{\delta\in\mathbb{D}}q_{s,\alpha}(Y_{B(\delta(w))})\geq
\max\{q_{s,\alpha}(Y_{0}),q_{s,\alpha}(Y_{1})\}.$ But given that $q_{s,\alpha
}(Y_{B(\delta(w))})\leq\max\{q_{s,\alpha}(Y_{0}),q_{s,\alpha}(Y_{1})\}$ for
any $\delta\in\mathbb{D}$ by part (ii), the claim follows. $\square$\pagebreak

\textbf{References}

\begin{description}
\item Aradillas Fern\'{a}ndez, A., J. Blanchet, J.L. Montiel Olea, C. Qiu, J.
Stoye, and L. Tan (2025a), \textquotedblleft$\varepsilon$-Minimax Solutions of
Statistical Decision Problems,\textquotedblright\ unpublished working paper,
Cornell University.

\item Berger, J. (1985), Statistical Decision Theory and Bayesian Analysis,
Second Edition, New York: SpringerVerlag.

\item Chen, H. and P. Guggenberger (2025), \textquotedblleft A note on minimax
regret rules with multiple treatments in finite samples,\textquotedblright%
\ forthcoming in \emph{Econometric Theory}.

\item Chambers, C. (2009), \textquotedblleft An axiomatization of quantiles on
the domain of distribution functions,\textquotedblright\ \emph{Mathematical
Finance} 19, 335--342.

\item Christensen, T., R. Moon, and F. Schorfheide (2023), \textquotedblleft
Optimal Discrete Decisions when Payoffs are Partially
Identified\textquotedblright, unpublished working paper.

\item Cucconi, O. (1968), \textquotedblleft Contributi all'Analisi Sequenziale
nel Controllo di Accettazione per Variabili. \emph{Atti dell' Associazione
Italiana per il Controllo della Qualit\`{a}} 6, 171--186.

\item De Castro, L., A. Galvao, and H. Ota (2026), \textquotedblleft Quantile
Approach to Intertemporal Consumption with Multiple Assets\textquotedblright,
\emph{Journal of Econometrics}, 253, 106161.

\item Guggenberger, P. and J. Huang (2025), \textquotedblleft On the numerical
approximation of minimax regret rules via fictitious play,\textquotedblright%
\ unpublished working paper, Pennsylvania State University.

\item Gupta, S. and S. Hande (1992), \textquotedblleft On some nonparametric
selection procedures. In: Saleh, A.K.Md.E. (Ed.), Nonparametric Statistics and
Related Topics. Elsevier.

\item Hirano, K. (2025), \textquotedblleft Waldean and Post-Waldean
Econometrics for Policy Analysis and Experimental Design,\textquotedblright%
\ in preparation for the ES monograph series.

\item Hirano, K. and J. Porter (2009), \textquotedblleft Asymptotics for
Statistical Treatment Rules,\textquotedblright\ \emph{Econometrica}, 77, 1683--1701.

\item Kitagawa, T., S. Lee, and C. Qiu (2024), \textquotedblleft Treatment
Choice with Nonlinear Regret,\textquotedblright\ forthcoming in
\emph{Biometrika}.

\item Kitagawa, T. and A. Tetenov (2018), \textquotedblleft Who Should be
Treated? Empirical Welfare Maximization Methods for Treatment
Choice,\textquotedblright\ \emph{Econometrica}, 86, 591--616.

\item Manski, C. (1988), \textquotedblleft Ordinal UtilityModels of Decision
Making Under Uncertainty,\textquotedblright\ \emph{Theory and Decision}, 25, 79--104.

\item \_\_\_\_\_\_\_\_(2004), \textquotedblleft Statistical Treatment Rules
for Heterogeneous Populations,\textquotedblright\ \emph{Econometrica}, 72, 221--246.

\item \_\_\_\_\_\_\_\_ and A. Tetenov (2007), \textquotedblleft Admissible
Treatment Rules for a Risk-averse Planner with Experimental Data on an
Innovation,\textquotedblright\ \emph{Journal of Statistical Planning and
Inference}, 137, 1998--2010.

\item \_\_\_\_\_\_\_\_(2023), \textquotedblleft Statistical Decision Theory
Respecting Stochastic Dominance,\textquotedblright\ \emph{The Japanese
Economic Review}, 74, 447--469.

\item Masten, M. (2023), \textquotedblleft Minimax-regret treatment rules with
many treatments,\textquotedblright\ \emph{The Japanese Economic Review}, 74, 501--537.

\item Montiel Olea, J.L, C. Qiu, and J. Stoye (2023), \textquotedblleft
Decision Theory for Treatment Choice with Partial
Identification,\textquotedblright\ forthcoming in \emph{Review of Economic
Studies}.

\item Qi, Z., Y. Cui, Y. Liu, and J.-S. Pang (2019), \textquotedblleft
Estimation of Individualized Decision Rules Based on An Optimized
Covariate-dependent Equivalent of Random Outcomes,\textquotedblright%
\ \emph{SIAM Journal on Optimization} 29 (3), 2337--2362

\item Qi, Z., J. Pang, and Y. Liu (2023), \textquotedblleft On Robustness of
Individualized Decision Rules,\textquotedblright\ \emph{Journal of the
American Statistical Association, }118 (543), 2143--2157.

\item Rostek, M.J. (2010), \textquotedblleft Quantile Maximization in Decision
Theory,\textquotedblright\ \emph{The Review of Economic Studies, }77, 339--371.

\item Schlag, K. (2003), \textquotedblleft How to minimize maximum regret
under repeated decision-making,\textquotedblright\ EUI working paper.

\item \_\_\_\_\_\_ (2006), \textquotedblleft ELEVEN - Tests needed for a
recommendation,\textquotedblright\ EUI working paper, ECO No. 2006/2.

\item Stoye, J. (2007), \textquotedblleft Minimax regret treatment choice with
incomplete data and many treatments,\textquotedblright\ \emph{Econometric
Theory}, 23(1), 190--199.

\item \_\_\_\_\_\_ (2009), \textquotedblleft Minimax Regret Treatment Choice
with Finite Samples,\textquotedblright\ \emph{Journal of Econometrics}, 151, 70--81.

\item \_\_\_\_\_\_ (2012), \textquotedblleft Minimax Regret Treatment Choice
with Covariates or with Limited Validity of Experiments,\textquotedblright%
\ \emph{Journal of Econometrics}, 166, 138--156.

\item Tetenov, A. (2012), \textquotedblleft Statistical Treatment Choice Based
on Asymmetric Minimax Regret Criteria,\textquotedblright\ \emph{Journal of
Econometrics}, 166, 157--165.

\item Wald, A. (1945), \textquotedblleft Statistical Decision Functions Which
Minimize the Maximum Risk,\textquotedblright\ \emph{Annals of Mathematics},
46(2), 265--280.

\item \_\_\_\_\_\_ (1947), \textquotedblleft Foundations of a General Theory
of Sequential Decision Functions,\textquotedblright\ \emph{Econometrica},
15(4), 279--313.

\item \_\_\_\_\_\_ (1950), Statistical Decision Functions, New York: Wiley.

\item Wang, L., Y. Zhou, R. Song, and B. Sherwood (2018) \textquotedblleft
Quantile-Optimal Treatment Regimes,\textquotedblright\ \emph{Journal of the
American Statistical Association}, 113(523), 1243--1254.

\item Yata, K. (2023), \textquotedblleft Optimal Decision Rules Under Partial
Identification,\textquotedblright\ unpublished working paper.\pagebreak
\end{description}

\textbf{TABLE\ I: Maximal and mean regret over all 18564 states of nature
}$s\in S^{E}(6,12)$\textbf{ for four different treatment rules}

$%
\begin{array}
[c]{ccccccccccccc}%
\text{\textbf{Case I)}} & \delta^{ES} & \delta^{1} & \delta^{.5} & \delta^{0}
& \delta^{ES} & \delta^{1} & \delta^{.5} & \delta^{0} & \delta^{ES} &
\delta^{1} & \delta^{.5} & \delta^{0}\\
\alpha & .1 & .1 & .1 & .1 & .5 & .5 & .5 & .5 & .9 & .9 & .9 & .9\\
q_{\alpha}(Y_{0})=.1 &  &  &  &  &  &  &  &  &  &  &  & \\
\max & .9 & \mathbf{.1} & .9 & .9 & .1 & \mathbf{.1} & .9 & .9 & .1 &
\mathbf{.1} & .9 & .9\\
\text{mean} & .08 & \mathbf{.04} & .14 & .1 & .01 & \mathbf{.01} & .37 & .37 &
0 & \mathbf{0} & .08 & .75\\
q_{\alpha}(Y_{0})=.5 &  &  &  &  &  &  &  &  &  &  &  & \\
\max & \mathbf{.5} & \mathbf{.5} & \mathbf{.5} & \mathbf{.5} & \mathbf{.5} &
\mathbf{.5} & \mathbf{.5} & \mathbf{.5} & \mathbf{.5} & \mathbf{.5} &
\mathbf{.5} & \mathbf{.5}\\
\text{mean} & \mathbf{.43} & \mathbf{.35} & \mathbf{.36} & \mathbf{0} &
\mathbf{.13} & \mathbf{.11} & \mathbf{.15} & \mathbf{.08} & \mathbf{.01} &
\mathbf{0} & \mathbf{.07} & \mathbf{.35}\\
q_{\alpha}(Y_{0})=.9 &  &  &  &  &  &  &  &  &  &  &  & \\
\max & .9 & .9 & .9 & \mathbf{.1} & .9 & .9 & .9 & \mathbf{.1} & .9 & .9 &
.9 & \mathbf{.1}\\
\text{mean} & .9 & .75 & .75 & \mathbf{0} & .8 & .44 & .37 & \mathbf{0} &
.33 & .1 & .11 & \mathbf{.04}\\
\text{\textbf{Case II)}} & \delta^{ES} & \delta^{1} & \delta^{.5} & \delta^{0}
& \delta^{ES} & \delta^{1} & \delta^{.5} & \delta^{0} & \delta^{ES} &
\delta^{1} & \delta^{.5} & \delta^{0}\\
\alpha & .1 & .1 & .1 & .1 & .5 & .5 & .5 & .5 & .9 & .9 & .9 & .9\\
q_{\alpha}(Y_{0})=.1 &  &  &  &  &  &  &  &  &  &  &  & \\
\max & .84 & \mathbf{.1} & .9 & .9 & .1 & \mathbf{.1} & .9 & .9 & 0 &
\mathbf{.1} & .9 & .9\\
\text{mean} & .05 & \mathbf{.04} & .14 & .1 & .01 & \mathbf{.01} & .37 & .37 &
0 & \mathbf{0} & 0.08 & .75\\
q_{\alpha}(Y_{0})=.5 &  &  &  &  &  &  &  &  &  &  &  & \\
\max & \mathbf{.5} & \mathbf{.5} & \mathbf{.5} & \mathbf{.5} & \mathbf{.5} &
\mathbf{.5} & \mathbf{.5} & \mathbf{.5} & \mathbf{.02} & \mathbf{.5} &
\mathbf{.5} & \mathbf{.5}\\
\text{mean} & \mathbf{.06} & \mathbf{.35} & \mathbf{.36} & \mathbf{0} &
\mathbf{.08} & \mathbf{.11} & \mathbf{.15} & \mathbf{.08} & \mathbf{0} &
\mathbf{0} & \mathbf{0.07} & \mathbf{.35}\\
q_{\alpha}(Y_{0})=.9 &  &  &  &  &  &  &  &  &  &  &  & \\
\max & .9 & .9 & .9 & \mathbf{.1} & .9 & .9 & .9 & \mathbf{.1} & .03 & .9 &
.9 & \mathbf{.1}\\
\text{mean} & 0 & .75 & .75 & \mathbf{0} & 0.04 & .44 & .37 & \mathbf{0} &
0.01 & .1 & .11 & \mathbf{.04}%
\end{array}
$

\pagebreak

\textbf{TABLE\ II: Proportion (in \%) of states }$s\in S^{E}(6,12)$\textbf{
for which regret for the empirical success rule }$\delta^{ES}$\textbf{ is
smaller than regret of non-data rules }$\delta\in\{\delta^{1},\delta
^{.5},\delta^{0}\}$

$%
\begin{array}
[c]{cccccccccc}%
\text{\textbf{Case I)}} & \delta^{1} & \delta^{.5} & \delta^{0} & \delta^{1} &
\delta^{.5} & \delta^{0} & \delta^{1} & \delta^{.5} & \delta^{0}\\
\alpha & .1 & .1 & .1 & .5 & .5 & .5 & .9 & .9 & .9\\
q_{\alpha}(Y_{0})=.1 &  &  &  &  &  &  &  &  & \\
Prop(R(\delta^{ES},s)<R(\delta,s)) & 0 & 33.3 & 33.3 & 0 & 92.5 & 92.5 & 0 &
33.3 & 100\\
Prop(R(\delta^{ES},s)\leq R(\delta,s)) & 76.5 & 100 & 56.9 & 97.5 & 97.5 &
95.0 & 100 & 100 & 100\\
q_{\alpha}(Y_{0})=.5 &  &  &  &  &  &  &  &  & \\
Prop(R(\delta^{ES},s)<R(\delta,s)) & 0 & .5 & .5 & 3.0 & 30.8 & 30.8 & 0 &
31.6 & 91.7\\
Prop(R(\delta^{ES},s)\leq R(\delta,s)) & 61.0 & 62.7 & 8.3 & 90.7 & 76.2 &
58.6 & 99.5 & 99.5 & 97.8\\
q_{\alpha}(Y_{0})=.9 &  &  &  &  &  &  &  &  & \\
Prop(R(\delta^{ES},s)<R(\delta,s)) & 0 & 0 & 0 & 0 & 2.5 & 2.5 & 0 & 16.2 &
43.1\\
Prop(R(\delta^{ES},s)\leq R(\delta,s)) & 43.3 & 43.4 & 0 & 21.0 & 11.7 & 2.5 &
66.7 & 66.7 & 43.1\\
\text{\textbf{Case II)}} & \delta^{1} & \delta^{.5} & \delta^{0} & \delta^{1}
& \delta^{.5} & \delta^{0} & \delta^{1} & \delta^{.5} & \delta^{0}\\
\alpha & .1 & .1 & .1 & .5 & .5 & .5 & .9 & .9 & .9\\
q_{\alpha}(Y_{0})=.1 &  &  &  &  &  &  &  &  & \\
Prop(R(\delta^{ES},s)<R(\delta,s)) & 16.2 & 73.0 & 73.0 & .3 & 95.3 & 95.3 &
0 & 33.4 & 100\\
Prop(R(\delta^{ES},s)\leq R(\delta,s)) & 76.5 & 100 & 73.0 & 100 & 97.5 &
95.3 & 100 & 100 & 100\\
q_{\alpha}(Y_{0})=.5 &  &  &  &  &  &  &  &  & \\
Prop(R(\delta^{ES},s)<R(\delta,s)) & 71.2 & 73.4 & 71.6 & 13.6 & 40.7 & 39.8 &
2.2 & 33.8 & 91.7\\
Prop(R(\delta^{ES},s)\leq R(\delta,s)) & 98.3 & 100 & 79.6 & 100 & 86.6 &
65.3 & 100 & 100 & 97.8\\
q_{\alpha}(Y_{0})=.9 &  &  &  &  &  &  &  &  & \\
Prop(R(\delta^{ES},s)<R(\delta,s)) & 99.5 & 99.6 & 99.6 & 84.0 & 82.1 & 75.6 &
56.9 & 73.0 & 43.1\\
Prop(R(\delta^{ES},s)\leq R(\delta,s)) & 100 & 100 & 99.6 & 100 & 95.7 &
75.6 & 100 & 100 & 43.1
\end{array}
$
\end{document}